\documentclass[runningheads]{llncs}

\usepackage[T1]{fontenc}
\usepackage{amsmath}
\usepackage{amssymb}
\usepackage{booktabs}
\usepackage{graphicx}
\usepackage{tikz}
\usetikzlibrary{decorations.pathreplacing,calc,arrows.meta,positioning,patterns}
\usepackage{pgfplots}
\pgfplotsset{compat=1.18}
\usepackage{xcolor}
\definecolor{oai}{RGB}{16,124,124}
\definecolor{ant}{RGB}{217,119,54}
\definecolor{goo}{RGB}{52,138,64}
\definecolor{dee}{RGB}{124,72,176}
\definecolor{xai}{RGB}{198,52,72}
\usepackage{etoolbox}
\usepackage{url}

\newtoggle{anon}
\togglefalse{anon}

\newcommand{\Reach}{\mathrm{Reach}}
\newcommand{\E}{\mathbb{E}}

\newtoggle{draftnotes}
\togglefalse{draftnotes}

\begin{document}

\title{When Is Delegated Play Truthful? Within-Range Regret and the Trilemma of Aligned Delegation}
\titlerunning{When Is Delegated Play Truthful?}

\iftoggle{anon}{%
  \author{Anonymous Author(s)}
  \authorrunning{Anonymous}
  \institute{Affiliation withheld for double-blind review}
}{%
  \author{Taksch Dube}
  \authorrunning{T. Dube}
  \institute{Kent State University, Kent, OH, USA}
  %
}

\maketitle

\begin{abstract}
Advertisers delegate bidding to autobidders; users delegate tasks to language-model agents. In each case a person describes what they want to an automated proxy. The proxy then acts in a mechanism on their behalf. This is the revelation principle run in production, and it forces a question the classical theory assumes away: when is it optimal to describe yourself honestly to your own proxy?

We show the answer turns on a single quantity, the proxy's \emph{within-range regret}. The most a principal can gain by misreporting equals the regret of the proxy's honest-report action, measured against the actions the principal could have steered it to take. Honest self-description is optimal exactly when the proxy already plays the best action it can reach, that is, when it is loyal to the principal (Theorem~\ref{thm:ttp}). The identity unifies a line of auction-specific results on autobidding incentives, and it identifies when the faithful-communication assumption behind language-model elicitation proxies (Huang et al.) is optimal for the user: exactly when the proxy is loyal.

The identity has a sharp consequence for the guardrails placed on proxies, from bid caps to the alignment layer of a language model. No guardrail can be at once \emph{binding} (it displaces the truthful action from the proxy's best reachable outcome), \emph{truthful} (honest reporting stays optimal), and \emph{capability-preserving} (that best outcome stays reachable through some report); any two preclude the third (Theorem~\ref{thm:trilemma}). A safety constraint that alters what a model does while leaving its best output reachable makes honest description of intent suboptimal, so a better-aimed report can gain. This is the incentive behind prompt-engineering and jailbreaking, though whether the exploiting report is easy to find is a separate question.

Because within-range regret is \#\textrm{P}-hard to compute exactly, we estimate it from samples and maintain it as a model is updated, at a cost set by how far the model drifts rather than how often it changes. Running that estimate on production language models from five providers under an alignment-style cap we impose, we find honest reporting leaves surplus unclaimed on every model, recovered by inflating the report.


\end{abstract}

\section{Introduction}
\label{sec:intro}

The revelation principle is the workhorse of mechanism design: whatever a strategic mechanism achieves in equilibrium, a \emph{direct} mechanism achieves by asking each participant to report its private type, under the promise that honest reporting is optimal \cite{gibbard1973,myerson1979,dasgupta1979,myerson1981}. The promise is constructive: given a mechanism and an equilibrium, one collects types, computes each agent's equilibrium message, and applies the original rule. But the construction quietly consumes four things a designer rarely has: the equilibrium itself, PPAD-complete to compute \cite{dgp2009,chendengteng2009,rubinstein2016}; knowledge of the prior and payoffs it depends on; a trusted party to run the composition; and a channel that carries a full type \cite{nisansegal2006}. For four decades the principle has been a reduction without an implementation.

Delegation now supplies all four, in production. An autobidder or a language-model agent is the executor that learns the equilibrium the designer cannot compute \cite{aggarwal2019,deng2021towards}; it carries the prior no one wrote down, fit to its environment from data; the platform that runs the mechanism is the trusted party; and the principal's report is the channel, a few honest sentences in place of a full contingent plan. We call the resulting object the \emph{delegation stack}: a principal describes its type to a \emph{proxy} $\pi_i \colon \Theta_i \to M_i$ that maps the report to a message, and principals face the \emph{wrapped} direct mechanism $g \circ \pi$. Delegation also adds one ingredient the classical construction never had: a proxy whose objective may not be the principal's. The premise that honest reporting is optimal is therefore no longer automatic, and becomes a question about a deployed system: when is it optimal to describe yourself honestly to your own proxy? Huang et al.\ \cite{huang2025elicitation}, building language-model elicitation proxies, adopt the truthful-reporting setting as their starting point and do not model the principal's incentive to misreport to its proxy. We analyze it.

\begin{figure}[t]
\centering
\begin{minipage}[c]{0.46\textwidth}
\centering
\begin{tikzpicture}[font=\footnotesize, >={Stealth[round]}]
  \node[draw, minimum width=0.88cm, minimum height=2.45cm] (g) at (3.25,0) {$g$};
  \foreach \i/\y in {1/1.0, 2/0, 3/-1.0} {
    \node[draw, circle, minimum size=0.52cm, inner sep=0] (t\i) at (0,\y) {$\theta_{\i}$};
    \node[draw, rounded corners, minimum width=0.72cm, minimum height=0.46cm] (q\i) at (1.6,\y) {$\pi_{\i}$};
    \draw[->] (t\i) -- (q\i);
    \draw[->] (q\i) -- (g.west|-q\i);
  }
  \draw[->] (g.east) -- ++(1.15,0);
  \node[above, font=\scriptsize] at (4.35,0) {outcome};
  \node[font=\scriptsize] at (0,-1.6) {principals};
  \node[font=\scriptsize] at (1.6,-1.6) {proxies};
  \node[font=\scriptsize] at (3.25,-1.6) {mechanism};
  \useasboundingbox (-0.45,-1.95) rectangle (4.9,1.95);
\end{tikzpicture}
\\[2pt] {\footnotesize (a) the delegation stack}
\end{minipage}\hfill
\begin{minipage}[c]{0.53\textwidth}
\centering
\begin{tikzpicture}[font=\footnotesize, lensnum/.style={circle,draw,fill=white,inner sep=0,minimum size=2.7mm,font=\tiny}]
  \def\rr{0.98}
  \fill[blue!45, fill opacity=0.32]   (90:0.54)  circle (\rr);
  \fill[green!45, fill opacity=0.32]  (210:0.54) circle (\rr);
  \fill[orange!55, fill opacity=0.32] (330:0.54) circle (\rr);
  \begin{scope}
    \clip (90:0.54) circle (\rr);
    \clip (210:0.54) circle (\rr);
    \clip (330:0.54) circle (\rr);
    \fill[pattern=north east lines, pattern color=red!75] (-1.5,-1.5) rectangle (1.5,1.5);
  \end{scope}
  \draw[blue!60!black]   (90:0.54)  circle (\rr);
  \draw[green!55!black]  (210:0.54) circle (\rr);
  \draw[orange!80!black] (330:0.54) circle (\rr);
  \node[above, font=\scriptsize] at (90:1.56) {capability-preserving};
  \node[below, font=\scriptsize] at (210:1.62) {binding};
  \node[below, font=\scriptsize] at (330:1.62) {truthful};
  \node[lensnum] at (150:0.66) {1};
  \node[lensnum] at (30:0.66) {2};
  \node[lensnum] at (270:0.64) {3};
  \begin{scope}[shift={(2.3,0)}]
    \node[lensnum] at (0,0.66) {1}; \node[anchor=west, font=\tiny] at (0.22,0.66) {output-clipping};
    \node[lensnum] at (0,0.22) {2}; \node[anchor=west, font=\tiny] at (0.22,0.22) {non-binding};
    \node[lensnum] at (0,-0.22) {3}; \node[anchor=west, font=\tiny] at (0.22,-0.22) {capability-losing};
    \fill[pattern=north east lines, pattern color=red!75] (-0.13,-0.79) rectangle (0.13,-0.53);
    \draw (-0.13,-0.79) rectangle (0.13,-0.53);
    \node[anchor=west, font=\tiny] at (0.22,-0.66) {none};
  \end{scope}
  \useasboundingbox (-1.9,-1.95) rectangle (4.45,1.95);
\end{tikzpicture}
\\[2pt] {\footnotesize (b) a guardrail is at most two of three}
\end{minipage}
\caption{(a) The delegation stack: each principal reports its type to a proxy, which messages the mechanism $g$, so principals face the wrapped mechanism $g \circ \pi$. (b) A guardrail composed onto a proxy cannot be binding, truthful, and capability-preserving at once (Theorem~\ref{thm:trilemma}); which pair a design keeps depends on its placement and its allowed set $D$ (Proposition~\ref{prop:placement}). Constraint-aware decoding stays truthful; an output-clip that leaves the optimum reachable ($a^\star \in D$) keeps capability but not truthfulness (the jailbreak corner, $W_i > 0$); a constraint that removes it ($a^\star \notin D$) can keep truthfulness, as the hard cap does, but not capability; a non-binding guardrail leaves the honest action already optimal.}
\label{fig:trilemma}
\end{figure}
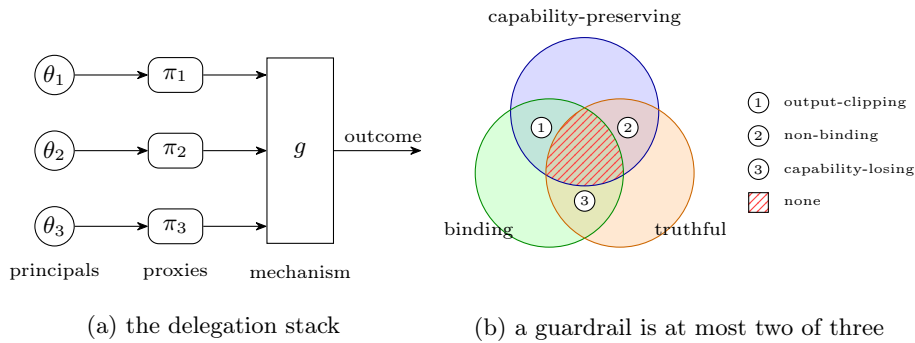

One identity answers it (Theorem~\ref{thm:ttp}): the most a principal gains by misreporting to its proxy equals the proxy's regret over the actions it can reach, its \emph{within-range regret}, which vanishes exactly when the proxy is loyal over that range and grows with its misalignment. This keeps two honesties distinct and binds them. The principal's \emph{truthful reporting} to its proxy, the question we ask, is optimal exactly when the proxy is \emph{loyal} over the range it can reach, that is, when it serves the principal rather than the platform. The identity's consequence is a limit on the guardrails placed on proxies, from bid caps to a language model's alignment layer (Theorem~\ref{thm:trilemma}; Fig.~\ref{fig:trilemma}). And because within-range regret is \#\textrm{P}-hard to compute exactly but estimable from samples, incentive compatibility becomes a property a platform measures rather than proves once.\footnote{Code and data reproducing every figure are provided as a reproducibility archive (Section~\ref{sec:empirics}).} For a language model, a binding safety layer that leaves the best output reachable rewards prompt-engineering.

\paragraph{Contributions.} One quantity runs through the paper; each result sharpens the last or puts it to work. (i)~A principal's gain from misreporting to its proxy equals the proxy's \emph{within-range regret} $W_i$ (Theorem~\ref{thm:ttp}), so honest reporting is optimal exactly when the proxy is loyal over the range it can reach. This unifies the first- and second-price autobidding incentive-compatibility results as one sign change in $W_i$ (Example~\ref{ex:autobid}). (ii)~Since $W_i$ alone governs misreporting, every guardrail faces a \emph{trilemma} (Theorem~\ref{thm:trilemma}): none is binding, truthful, and capability-preserving at once. Its placement decides whether truthfulness is guaranteed and its allowed set whether capability survives (Proposition~\ref{prop:placement}), and a conservation law bounds the loss (Proposition~\ref{prop:quant-trilemma}). (iii)~Computing $W_i$ exactly is \#\textrm{P}-hard (Proposition~\ref{prop:hardness}), but the identity turns certifying a wrapped mechanism into \emph{estimating $W_i$ from samples} of the proxy, an estimate that stays valid as the proxy drifts (Proposition~\ref{prop:online}). (iv)~We run that sampling estimator on production language models from five providers: under an alignment-style cap we impose, sampling each proxy's play recovers $W_i > 0$ with report inflation above one on every model, so the certification of~(iii) runs on deployed models and reproduces the incentive (i)--(ii) predict.

Section~\ref{sec:related} places the results in the autobidding, delegation, and incentive-estimation literatures; the rest sets up the delegation stack and within-range regret (Section~\ref{sec:model}), bounds and then exactly characterizes the gain from misreporting (Section~\ref{sec:inputgame}), turns that into the trilemma (Section~\ref{sec:limits}), makes incentive compatibility a measured, maintainable quantity (Section~\ref{sec:certification}), measures within-range regret on production language models (Section~\ref{sec:empirics}), and discusses scope and open problems (Section~\ref{sec:discussion}). One assumption holds throughout: the platform commits to running the published $g \circ \pi$, the standard commitment of mechanism design. We ask nothing else of a proxy beyond being a fixed map; where we need it to play well for its principal, we say so. We do not assume the proxy is loyal. A platform-operated proxy may answer to the platform, so our results describe the principal's incentive against whatever proxy it faces; certifying loyalty remains open. The formal results are theorems about a Bayesian game with proxies; reading a guardrail as an alignment layer is an interpretation, one we support by measuring within-range regret on deployed models.


\section{Related Work}
\label{sec:related}

\paragraph{Delegation, learning agents, and autobidding.}
The revelation principle and its failure regimes are well charted, including limited commitment, reporting costs, and sequential rationality \cite{bester2001,dovalskreta2022,kephartconitzer2016,conitzersandholm2004,sugaya2021}; we work where it is valid and supply its inputs by delegation, a device with a long pedigree: strategic commitment \cite{fershtmanjudd1987}, communication-equilibrium mediators \cite{forges1986,myerson1986}, program equilibrium \cite{tennenholtz2004}, and proxy bidding \cite{ausubelmilgrom2002,ashlagi2009mediators}. Closest are Kolumbus and Nisan's user-of-learning-agents models \cite{kolumbusnisan2022manipulate,kolumbusnisan2022auctions}: a principal delegating to a no-regret agent can gain by misreporting to it, and their first-versus-second-price reversal is, by Theorem~\ref{thm:ttp}, the within-range-regret identity made auction-specific; we give the closed-form gain at a fixed proxy. The designer-side mirror, a mechanism facing a no-regret agent rather than a principal steering its own, is charted separately \cite{braverman2018selling,deng2019strategizing,camara2020mechanisms}; our object is the principal's incentive to its proxy, not the designer's leverage over it. What is new is not a theorem about manipulation gain but the separation the identity makes visible, between how well a proxy plays the world ($R_i$) and how well its input parameterization is aimed ($W_i$), together with the trilemma and certification that separation supports; the identity itself generalizes and renames their reversal, and we claim no more for it. The autobidding literature is the deployed special case \cite{aggarwal2019,deng2021towards,deng2022nontruthful,nonuniform2023,raic2025}: whether an advertiser gains by misreporting its constraints to its own autobidder is autobidding incentive compatibility, which fails for both first- and second-price rules \cite{alimohammadi2023aic} and makes single-round incentive compatibility the wrong concept once advertisers only set proxy objectives \cite{litang2024}; such a failure is exactly $W_i > 0$ in one format, which Theorem~\ref{thm:ttp} locates.

\paragraph{Language-model agents and certification from data.}
Machine agents increasingly transact in markets, as Parkes and Wellman \cite{parkeswellman2015} anticipated: language models act as auction participants \cite{shah2025synthetic}, bidders in token auctions \cite{duetting2024llm}, agents in matching markets \cite{hoshino2026matching}, and elicitation proxies that take communication to be faithful by construction \cite{huang2025elicitation}, which their model does not analyze. On certification, deciding equilibrium is intractable \cite{dgp2009,chendengteng2009,rubinstein2016}, so within-range regret is estimated: approximate incentive compatibility is learnable from samples \cite{balcan2019} and has been verified for auction networks and estimated from production logs \cite{curry2020certifying,deng2020datadriven,pieroth2024verifying}. We certify the within-range quantity specifically, maintain it under drift, and calibrate against the known underestimation of deep-learning regret estimators \cite{you2026regret}.

\paragraph{Impossibility trilemmas and constrained agents.}
Our trilemma joins a family of mechanism-design impossibilities in which three pairwise-attainable properties cannot hold together. The nearest is the credible-auction trilemma of Akbarpour and Li \cite{akbarpourli2020}, where no optimal auction is at once credible, strategy-proof, and static; theirs is about an operator who might deviate undetectably, ours about what a range restriction does to a delegated proxy's reachable optimum. The two-way form of our binding-versus-capability tension is already visible elsewhere: a binding safety constraint forfeits capability in fine-tuning \cite{chen2026safety}, strategy-proofness costs a bounded multiple of policy value in RLHF \cite{kleinebuening2025rlhf}, and robustness to gaming costs accuracy in strategic classification \cite{hardt2016strategic}; the trilemma adds the truthfulness axis they omit.


\section{Preliminaries}
\label{sec:model}

\paragraph{Environment.}
There are $n$ agents. Agent $i$ has a \emph{type} $\theta_i \in \Theta_i$, the private information only it knows, such as its value for an item or what it actually wants. The type profile is drawn from a common prior $F$ on $\Theta = \prod_i \Theta_i$. An outcome $x \in X$ gives agent $i$ utility $u_i(x, \theta_i) \in [0, H]$ (private values). Type and message sets are finite for exposition; the arguments extend to continuum types and actions wherever the maxima invoked are attained.

\paragraph{Mechanisms and regret.}
A mechanism is $\Gamma = (M, g)$ with \emph{message sets} $M_i$, the actions an agent can put into the mechanism such as a bid, and an outcome rule $g: M \to X$ that maps the messages to an outcome. A strategy is $s_i: \Theta_i \to M_i$, a rule for turning a type into a message. Interim utility under profile $s$ is
\[
U_i(s \mid \theta_i) \;=\; \E_{\theta_{-i} \sim F(\cdot \mid \theta_i)}\!\left[ u_i\big(g(s_i(\theta_i), s_{-i}(\theta_{-i})), \theta_i\big) \right],
\]
and $U_i(m, s_{-i} \mid \theta_i)$ denotes the same expectation with the fixed message $m$ in place of $s_i(\theta_i)$. The \emph{interim regret} of $s$ for agent $i$ is
\[
R_i(s) \;=\; \max_{\theta_i \in \Theta_i} \left[ \max_{m \in M_i} U_i(m, s_{-i} \mid \theta_i) \;-\; U_i(s \mid \theta_i) \right].
\]
The profile $s$ is an interim $\varepsilon$-Bayes-Nash equilibrium ($\varepsilon$-BNE) if $\max_i R_i(s) \le \varepsilon$. A mechanism with $M_i = \Theta_i$ is \emph{direct}; it is interim $\varepsilon$-incentive compatible ($\varepsilon$-IC) if the truthful profile is an interim $\varepsilon$-BNE of it~\cite{dgv1979,myerson1981}, the additive-$\varepsilon$ form being the standard approximate relaxation.

\paragraph{The delegation stack.}
A \emph{proxy} for agent $i$ is a map $\pi_i: \Theta_i \to M_i$ from input reports to messages, and $\pi = (\pi_1, \dots, \pi_n)$ is the \emph{proxy profile}. The \emph{wrapped mechanism} is the direct mechanism
\[
\Gamma^{\pi} \;=\; \big(\Theta,\; g \circ \pi\big), \qquad (g \circ \pi)(t) = g\big(\pi_1(t_1), \dots, \pi_n(t_n)\big).
\]
The Bayesian game induced by $\Gamma^{\pi}$ among the principals is the \emph{input game}: each principal knows $\theta_i$ and chooses a report $t_i$. The \emph{reachable set} of proxy $i$ is $\Reach_i = \pi_i(\Theta_i) \subseteq M_i$. The report language is the type space itself, so $\Reach_i$ is exactly the set of messages a type-report can steer. A richer or coarser language would change $\Reach_i$, and with it the within-range regret of Section~\ref{sec:inputgame}.

Proxies are deterministic maps; randomized proxies replace messages with distributions and every statement holds in expectation. How $\pi$ is produced is irrelevant to Section~\ref{sec:inputgame} and becomes the subject of Section~\ref{sec:certification}.

\paragraph{Commitment.}
As stated in the introduction, the platform commits to running the published $g \circ \pi$ and each proxy is a fixed map. The one place that additionally needs a proxy to play well for its principal is the inheritance statement of Section~\ref{sec:erp}, where we name the assumption.


\section{Within-Range Regret and the Truth-to-Proxy Identity}
\label{sec:inputgame}

A principal's only lever is the report it gives its proxy. We first bound what that lever is worth: the wrapped mechanism inherits incentive compatibility from the proxy's equilibrium quality. We then characterize that gain exactly as the proxy's \emph{within-range regret}, reducing strategic self-description to a geometric property of the proxy's range. Each result is stated here and proved in full in Appendix~\ref{app:proofs}.

\subsection{Inheritance}
\label{sec:erp}

\begin{proposition}[Inheritance]
\label{prop:inherit}
If the proxy profile $\pi$ is an interim $\varepsilon$-BNE of $\Gamma$, then the wrapped mechanism $\Gamma^{\pi}$ is interim $\varepsilon$-IC, and under truthful reports its outcome coincides with the play of $\pi$ in $\Gamma$ at every type profile.
\end{proposition}

The classical revelation principle is the $\varepsilon = 0$ case. Wrapping is conservative: it never enlarges the strategic surface ($\Reach_i \subseteq M_i$) and adds no instrument absent from the base mechanism, in particular no Cr\'emer--McLean surplus extraction \cite{cremermclean1988}.

How small can $\varepsilon$ be? Not zero in general: an exact equilibrium ($\varepsilon = 0$), or even an inverse-polynomial approximation, is PPAD-hard to compute \cite{dgp2009,chendengteng2009,rubinstein2016}, already under complete information. The slack is a property of the problem, so the guarantee must be \emph{measured} rather than assumed. This is what makes the certification of Section~\ref{sec:certification} necessary. Where an $\varepsilon$-equilibrium is efficiently computable, in zero-sum and potential games \cite{cesabianchi2006} or no-regret-learnable auctions \cite{kolumbusnisan2022auctions}, the designer knows $\varepsilon$ outright. Otherwise the profile comes from outside, an autobidder \cite{aggarwal2019} or a fine-tuned model \cite{huang2025elicitation}, and its $\varepsilon$ is learned by sampling.


\subsection{Coarsening and the Identity}

Recall the \emph{input game} from Section~\ref{sec:model}: the game the principals play when each chooses a report to hand its proxy. Each principal can drive its proxy only to actions in $\Reach_i$, so delegation restricts the base game to the strategies whose range lies in $\Reach_i$. This restriction of the strategy space is what we call \emph{coarsening}. The next lemma identifies the input game with the coarsened base game exactly.

\begin{lemma}[The input game is the coarsened base game]
\label{lem:coarsen}
Fix $\pi$. For each $i$, the map $\tau_i \mapsto \pi_i \circ \tau_i$ takes report strategies $\tau_i: \Theta_i \to \Theta_i$ of the input game onto exactly the base-game strategies with range in $\Reach_i$, and it preserves interim utilities. Consequently the equilibria of the input game correspond to the equilibria of $\Gamma$ with strategy sets restricted to $\Reach_i$-valued maps.
\end{lemma}

Delegation is strategy-space coarsening. The manipulation identity and the guardrail corollary that follow are both corollaries of where the coarsened set sits relative to best responses.

\begin{definition}[Within-range regret and manipulation gain]
For a true type $\theta_i$, define
\begin{align*}
W_i(\theta_i; \pi) &= \max_{m \in \Reach_i} U_i(m, \pi_{-i} \mid \theta_i) \;-\; U_i\big(\pi_i(\theta_i), \pi_{-i} \mid \theta_i\big),\\[2pt]
G_i(\theta_i; \pi) &= \max_{t_i \in \Theta_i} U_i\big(\pi_i(t_i), \pi_{-i} \mid \theta_i\big) \;-\; U_i\big(\pi_i(\theta_i), \pi_{-i} \mid \theta_i\big).
\end{align*}
The first is the proxy's \emph{within-range regret} at $\theta_i$; the second is the principal's \emph{manipulation gain}, the most it can gain by misreporting its own type to the proxy. The proxy is \emph{range-optimal} at $\theta_i$ if $W_i(\theta_i; \pi) = 0$.
\end{definition}

\begin{theorem}[Truth-to-proxy]
\label{thm:ttp}
For every $i$ and $\theta_i$:
(i) $G_i(\theta_i; \pi) = W_i(\theta_i; \pi)$;
(ii) $G_i(\theta_i; \pi) \le R_i(\pi)$;
(iii) truthful reporting is interim optimal at $\theta_i$ if and only if $\pi_i$ is range-optimal at $\theta_i$, and $\Gamma^{\pi}$ is exactly IC if and only if $\pi$ is range-optimal at every type of every agent.
\end{theorem}

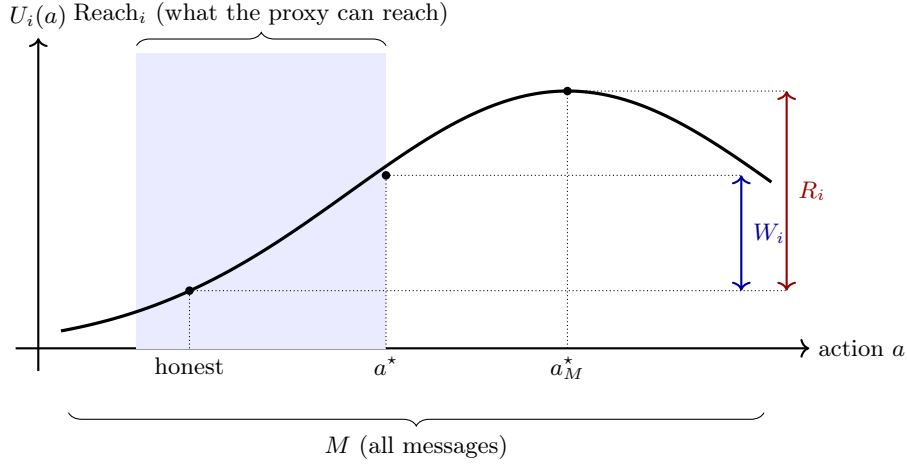
\begin{figure}[tbp]
\centering
\begin{tikzpicture}[scale=1.0,every node/.style={font=\small}]
  \draw[->,thick] (-0.3,0) -- (10.2,0) node[right] {action $a$};
  \draw[->,thick] (0,-0.3) -- (0,4.1) node[above] {$U_i(a)$};
  \fill[blue!8] (1.3,0) rectangle (4.6,3.9);
  \draw[very thick,domain=0.3:9.7,samples=120,smooth] plot (\x,{3.4*exp(-0.06*(\x-7)*(\x-7))});
  \def\xh{2.0} \def\xwr{4.6} \def\xr{7.0} \def\uh{0.762} \def\uwr{2.286} \def\ur{3.4}
  \foreach \x/\u in {\xh/\uh,\xwr/\uwr,\xr/\ur}{ \fill (\x,\u) circle (1.6pt); \draw[densely dotted] (\x,0) -- (\x,\u); }
  \node[below] at (\xh,0) {\footnotesize honest};
  \node[below] at (\xwr,0) {\footnotesize $a^\star$};
  \node[below] at (\xr,0) {\footnotesize $a^\star_M$};
  \draw[decorate,decoration={brace,amplitude=5pt,mirror}] (0.4,-0.85) -- (9.6,-0.85) node[midway,below=5pt] {\footnotesize $M$ (all messages)};
  \draw[decorate,decoration={brace,amplitude=5pt}] (1.3,4.0) -- (4.6,4.0) node[midway,above=4pt] {\footnotesize $\Reach_i$ (what the proxy can reach)};
  \draw[<->,thick,blue!60!black] (9.3,\uh) -- (9.3,\uwr) node[midway,right=1pt] {\footnotesize $W_i$};
  \draw[<->,thick,red!60!black]  (9.9,\uh) -- (9.9,\ur)  node[midway,right=1pt] {\footnotesize $R_i$};
  \draw[densely dotted] (\xh,\uh) -- (9.9,\uh);
  \draw[densely dotted] (\xwr,\uwr) -- (9.3,\uwr);
  \draw[densely dotted] (\xr,\ur) -- (9.9,\ur);
\end{tikzpicture}
\caption{Within-range regret is what governs misreporting. The proxy reaches only $\Reach_i \subseteq M_i$, so a principal's gain from misreporting equals the gap $W_i$ to the best \emph{reachable} action, not the unrestricted regret $R_i$ to the global optimum $a^\star_M$ (Theorem~\ref{thm:ttp}).}
\label{fig:reachable}
\end{figure}

The theorem separates two quality metrics that deployment conflates: the unrestricted regret $R_i$, how well the proxy plays against the world, and the within-range regret $W_i$, how well its input parameterization is aimed. Here $W_i \le R_i$, and $W_i$ alone governs the temptation to lie. A constant proxy has $W_i = 0$ but is useless, so the design problem is $W_i = 0$ with small $R_i$. The autobidding incentive-compatibility results (Section~\ref{sec:related}) are instances of the identity: the truthful rule makes the honest report range-optimal and the others do not. Figure~\ref{fig:reachable} draws the geometry, and Example~\ref{ex:autobid} works out the canonical pair.

The geometry takes a closed form in the canonical auction case, recovering the autobidding results as a sign change in within-range regret.

\begin{example}[The first-price/second-price reversal is a sign change in $W_i$]
\label{ex:autobid}
An advertiser of value $v$ delegates to a value-faithful autobidder, the proxy $\pi(t) = t$ that bids the reported value, against a competing bid $c \sim F$. Every bid is reachable, $\Reach = [0, H]$, so Theorem~\ref{thm:ttp} reads the incentive to misreport off the interim payoff in each format. Under a \emph{second-price} rule the payoff of bid $b$ is $\int_0^b (v - c)\,dF(c)$, maximized at $b = v$: the honest report is range-optimal, $W_i(v) = 0$, and autobidding IC holds. Under a \emph{first-price} rule the payoff is $(v - b)\,F(b)$, maximized at the bid $b^\star < v$ solving $(v - b^\star) f(b^\star) = F(b^\star)$; the honest bid $v$ earns $(v-v)F(v) = 0$, so
\[
W_i(v) \;=\; (v - b^\star)\,F(b^\star) \;>\; 0,
\]
and the advertiser strictly gains by reporting $b^\star$ in place of $v$. Honest self-description is optimal exactly in the format whose optimal bid is the honestly reported one. For this value-bidding proxy that format is the second-price auction; for the regret-minimizing agents of Kolumbus and Nisan \cite{kolumbusnisan2022manipulate,kolumbusnisan2022auctions} it is the first-price auction. The sign of $W_i$ follows the proxy, not the format. The autobidding incentive-compatibility failures of \cite{alimohammadi2023aic} are instances of the same identity.
\end{example}

The within-range restriction is not vacuous: across $60{,}000$ random Bayesian games $W_i$ falls strictly below $R_i$ in $58\%$ of instances (mean $R_i = 1.64$ against $W_i = 1.17$), so a delegation analysis must track $W_i$, not the unrestricted regret a naive bound would use. This computation uses \iftoggle{anon}{a market-simulation library}{the \texttt{llm-market-sim} library}~\cite{llmmarketsim}; the full distribution is Figure~\ref{fig:rwgap} in Appendix~\ref{app:empirics}, and the language-model study is Section~\ref{sec:empirics}.

\subsection{Guardrails and the Input Channel}
\label{sec:guardrails}

A \emph{guardrail} is a constraint composed onto the proxy that can only narrow its reach: $\pi'_i = c \circ \pi_i$ for some $c: M_i \to M_i$ with $c(\Reach_i) \subseteq \Reach_i$ (a bid cap, a prudence rule, a filtered action set, a policy-constrained decoder), applied to agent $i$ with the other proxies held fixed, so within-range regret is measured against the opponents $\pi_{-i}$. The condition $c(\Reach_i) \subseteq \Reach_i$ says a guardrail removes reachable actions rather than synthesizing new ones. It holds exactly for the interval reach of a bid-scaling proxy, and up to discretization for the sampled report grid of Section~\ref{sec:empirics}. Every guardrail in this paper meets it.

\begin{corollary}[Guardrails]
\label{cor:guardrails}
If under the guardrailed profile $\pi'$ the constrained action $\pi'_i(\theta_i)$ is not optimal within the constrained range $\Reach'_i$ at $\theta_i$, then $G_i(\theta_i; \pi') = W_i(\theta_i; \pi') > 0$: the guardrail makes truthful self-description strictly suboptimal.
\end{corollary}

The corollary is Theorem~\ref{thm:ttp} pointed at a design practice. A prudence rule caps bids at forty percent of the reported value, where the best response is half the true value, so the principal inflates its report by $1.25\times$ to restore the intended bid. The guardrail changed no feasible outcome, only the semantics of the input channel. The prediction is falsifiable: inflation concentrates on the principals for whom the constraint binds.

\section{The Trilemma of Aligned Delegation}
\label{sec:limits}

Corollary~\ref{cor:guardrails} says a guardrail that displaces the optimal reachable action breaks truth-to-proxy. This is not an incidental side effect but a structural limit, and it answers, at the level of the proxy, the question in the title: a guardrail that does real work cannot also keep self-description honest without forfeiting the principal's best outcome.

Fix a type $\theta_i$ and let $a^\star = \arg\max_{m \in \Reach_i} U_i(m, \pi_{-i} \mid \theta_i)$ be the honest-optimal reachable action, assumed unique. A guardrail $c$ with constrained range $\Reach'_i = c(\Reach_i) \subseteq \Reach_i$ has three properties of interest at $\theta_i$:
\begin{description}
\item[Binding:] $c(\pi_i(\theta_i)) \neq a^\star$, so the guardrailed truthful action differs from the honest optimum.
\item[Truthful:] $W_i(\theta_i; \pi') = 0$, so honest reporting is optimal within $\Reach'_i$.
\item[Capability-preserving:] $a^\star \in \Reach'_i$, so the honest-optimal action remains reachable through some report.
\end{description}

Here \emph{binding} is a property of the composite $c \circ \pi_i$, not of $c$ alone: a do-nothing guardrail on a proxy that already plays below $a^\star$ still counts as binding, since the test is whether the honest action sits at the optimum, not whether $c$ moved it.

\begin{theorem}[Aligned delegation is a trilemma]
\label{thm:trilemma}
At any type with a unique honest-optimal reachable action, no guardrail is simultaneously binding, truthful, and capability-preserving. Any two of the three preclude the third.
\end{theorem}

The impossibility is qualitative and assumes a unique optimum. It has a quantitative form that drops uniqueness, covers randomized proxies, and turns the three-way conflict into a rate. Equip each $M_i$ with a metric $d_i$, and call the optimum $(\varepsilon, \delta)$-\emph{sharp} at $\theta_i$ if every reachable action at distance at least $\varepsilon$ from $a^\star$ loses at least $\delta$: $U_i(m, \pi_{-i} \mid \theta_i) \le U^\star - \delta$ whenever $m \in \Reach_i$ and $d_i(m, a^\star) \ge \varepsilon$, where $U^\star = \max_{m \in \Reach_i} U_i(m, \pi_{-i} \mid \theta_i)$.

\begin{proposition}[Conservation of the binding loss]
\label{prop:quant-trilemma}
Suppose the optimum is $(\varepsilon, \delta)$-sharp at $\theta_i$, and let a guardrail be range-restricting ($\Reach'_i \subseteq \Reach_i$) and $\varepsilon$-binding, meaning $d_i(c(\pi_i(\theta_i)), a^\star) \ge \varepsilon$. Write its \emph{capability loss} $\kappa = U^\star - \max_{m \in \Reach'_i} U_i(m, \pi_{-i} \mid \theta_i) \ge 0$, zero exactly when the best reachable value is retained. Then the capability loss and the within-range regret obey a conservation law,
\[
\kappa \;+\; W_i(\theta_i; \pi') \;\ge\; \delta ;
\]
in particular a capability-preserving guardrail ($\kappa = 0$) has $W_i \ge \delta$.
\end{proposition}

Binding by $\varepsilon$ therefore imposes a loss budget $\delta$ that no design can avoid, only allocate: it is paid in capability ($\kappa$), in truthfulness ($W_i$), or split between them. The placements of Proposition~\ref{prop:placement} are the extreme allocations. Constraint-aware decoding spends the budget entirely on capability ($W_i = 0$, $\kappa \ge \delta$). A \emph{capability-preserving} output-clip spends it entirely on truthfulness ($\kappa = 0$, $W_i \ge \delta$); a \emph{capability-removing} one, such as a hard cap, can spend it on capability instead ($W_i = 0$, $\kappa \ge \delta$). The design chooses the currency, not the total. Sharpness sets the budget: a $\mu$-strongly-concave interim utility on a convex message set is $(\varepsilon, \tfrac{\mu}{2}\varepsilon^2)$-sharp, so $\delta$ grows as $\tfrac{\mu}{2}\varepsilon^2$. Because the condition is stated around the optimal \emph{value}, it applies to randomized proxies without change, with the binding distance read on the honest action almost surely.

For a language-model proxy the guardrail is the alignment layer and $\Reach_i$ is what the prompt can elicit. Theorem~\ref{thm:trilemma} then reads: an alignment layer that changes the output while keeping the genuinely best output promptable makes honest description of intent suboptimal, the incentive to prompt-engineer or jailbreak \cite{wei2023jailbroken,zou2023universal}.

\paragraph{Placement and the allowed set decide the corner.}
Which of the three a design surrenders splits along two axes: its placement decides whether truthfulness is guaranteed, and its allowed set whether capability survives. Call a proxy \emph{loyal} if its output maximizes the principal's own interim utility over the messages it can produce.

\begin{lemma}[Loyalty]
\label{lem:loyalty}
Suppose that for some action class $C_i \subseteq M_i$ the proxy selects an interim-optimal action within $C_i$ at each reported type,
\[
\pi_i(t_i) \in \arg\max_{m \in C_i} U_i(m, \pi_{-i} \mid t_i)
\qquad (\text{so } \pi_i(t_i) \in C_i).
\]
Then $W_i(\theta_i; \pi) = 0$ at every type of agent $i$, so truthful reporting is interim optimal for agent $i$. The argument uses no structure of $g$: for any mechanism whose interim utilities $U_i$ make the proxy loyal in this sense, truthfulness follows, and if every proxy is loyal the wrapped mechanism is incentive-compatible.
\end{lemma}

Within-range regret thus measures not the strength of a constraint but the proxy's \emph{misalignment}: it is positive only when the proxy optimizes something other than the principal's utility over its reachable set. A guardrail can keep the proxy loyal or break that loyalty, depending on where it acts.

\begin{proposition}[Placement]
\label{prop:placement}
Fix $\theta_i$ with a unique honest-optimal reachable action $a^\star$ and a constraint with allowed set $D \subseteq \Reach_i$. If the constraint is non-binding, a loyal proxy's honest output stays at $a^\star$, at once truthful and capability-preserving. A binding constraint cannot be both (Theorem~\ref{thm:trilemma}), and its placement fixes which it keeps. \emph{Constraint-aware decoding} replaces the proxy by one that re-optimizes within the allowed set, $\pi'_i(t_i) \in \arg\max_{m \in D} U_i(m, \pi_{-i} \mid t_i)$; by Lemma~\ref{lem:loyalty} with $C_i = D$ its honest output is optimal within $D$, so it stays \emph{truthful}, forfeiting capability exactly when $a^\star \notin D$. \emph{Output-clipping} instead composes a map onto the proxy output, $\pi'_i = c \circ \pi_i$ with $c(\Reach_i) = D$ (the guardrail of Theorem~\ref{thm:trilemma}), applied after the proxy has optimized. When it is capability-preserving ($a^\star \in D$) a binding clip displaces the honest output from the unique constrained optimum $a^\star$ and so forfeits \emph{truthfulness}; when $a^\star \notin D$ it may instead stay truthful, as the hard cap of Section~\ref{sec:empirics} does by clipping the honest output onto the constrained frontier, spending its budget on capability. Either way the binding, truthful, capability-preserving corner stays out of reach, which re-proves Theorem~\ref{thm:trilemma}.
\end{proposition}

When the optimum stays reachable ($a^\star \in D$), the same constraint occupies different corners according to where it is placed. For a language-model proxy the distinction is operational. A safety constraint built into decoding, where the model selects the best \emph{allowed} output, keeps honest prompting optimal while putting some outputs out of reach. A filter applied to the model's chosen output keeps every output reachable through a rephrased prompt, and so rewards prompt-engineering. Constraint-aware decoding and output filtering are not interchangeable.

\paragraph{From the individual to the equilibrium.}
Corollary~\ref{cor:guardrails}, the trilemma, and Proposition~\ref{prop:placement} hold the other proxies at their honest messages; they are single-agent statements. What a capability-preserving guardrail does to the \emph{equilibrium} splits on whether the mechanism couples the agents, and a structural fact ties the question to loyalty. If the proxy is loyal at every type, its honest output already maximizes over its reach, so $a^\star_i(\theta_i) = \pi_i(\theta_i)$ at every type and $\{a^\star_i(\theta_i)\}_{\theta_i}$ exhausts $\Reach_i$. Capability-preservation then forces $\Reach'_i = \Reach_i$, so $c$ maps the finite reachable set onto itself. A surjection of a finite set that is \emph{deflationary} ($c(m) \le m$, as every cap and prudence rule on the ordered bid space is) preserves the total of its arguments and is therefore the identity, so $c$ fixes every reachable action and nothing binds (idempotence yields the same conclusion on unordered messages). A binding, range-restricting, capability-preserving guardrail that is deflationary (or, on unordered messages, idempotent) therefore exists only on a \emph{non-loyal} proxy, and the equilibrium effects below are effects of that misalignment.

\begin{proposition}[Dominant-strategy neutrality]
\label{prop:neutral}
Suppose each agent has a unique weakly-dominant reachable action $a^\star_i(\theta_i)$: it maximizes $U_i(\cdot\,, m_{-i} \mid \theta_i)$ over $\Reach_i$ for every reachable opponent profile $m_{-i}$. If a range-restricting guardrail is capability-preserving ($a^\star_i(\theta_i) \in \Reach'_i$ at every type), then $(a^\star_i(\theta_i))_i$ is a weakly-dominant-strategy equilibrium of both the guarded and the unguarded input game, unique in the unguarded game and, in the guarded game, still weakly dominant though the smaller reach $\Reach'_{-i}$ may admit ties, and the two induce the same allocation, payments, revenue, and welfare; only the report realizing $a^\star_i$ changes, inflating for a monotone cap. Second-price and VCG composed with a value-faithful proxy are of this form.
\end{proposition}

\begin{proposition}[A capability-preserving guardrail can destroy welfare]
\label{prop:destroy}
Without dominance, capability-preservation does not lift to the equilibrium: there is a symmetric two-agent mechanism, a type-faithful proxy, and a range-restricting guardrail capability-preserving at every type, yet an equilibrium of the unguarded input game has no equal-outcome counterpart in the guarded game and strictly higher social welfare than every guarded equilibrium.
\end{proposition}

The witness is a stag hunt on reachable messages $\{\mathsf{stag}, \mathsf{hare}\}$ with $u(\mathsf{stag},\mathsf{stag}) = 4$, $u(\mathsf{stag},\mathsf{hare}) = 0$, $u(\mathsf{hare},\cdot) = 3$, and a type-faithful proxy. Against the honest opponent the best reply is $\mathsf{hare}$ at every type, so $a^\star_i = \mathsf{hare}$, and capability-preservation protects $\mathsf{hare}$, not $\mathsf{stag}$. Since $c(\pi_i(\theta_i)) = \mathsf{hare} = a^\star_i$ at every type, the guardrail is even non-binding in the single-agent sense of Section~\ref{sec:limits}: welfare destruction needs no binding at all, only the deletion of an off-equilibrium message. The efficient equilibrium $(\mathsf{stag},\mathsf{stag})$, welfare $8$, rests on $\mathsf{stag}$, a best reply only to $\mathsf{stag}$ and never to the honest opponent. A guardrail mapping $\mathsf{stag} \mapsto \mathsf{hare}$ is capability-preserving yet deletes the message that equilibrium needs, leaving the guarded game to play only $(\mathsf{hare},\mathsf{hare})$, welfare $6$. This is the strategic complementarity the single-agent analysis cannot see: capability-preservation protects each agent's best reply to \emph{honest} opponents, while a coordination equilibrium rests on best replies to \emph{each other}. A capability-preserving guardrail is outcome-neutral where the mechanism is already dominant-strategy, and can destroy efficient equilibria where it is not.

\paragraph{Measured on language-model proxies.}
Section~\ref{sec:empirics} runs this corner on production language models from five providers: sampling each proxy's play under an alignment-style cap, the estimator of Section~\ref{sec:certification}, recovers the predicted $W_i > 0$ with report inflation on every model, and a placement ablation separates the two corners of Proposition~\ref{prop:placement}.


\section{Certification}
\label{sec:certification}

The inheritance proposition (Proposition~\ref{prop:inherit}) and the hardness floor of Section~\ref{sec:erp} jointly imply that a wrapped mechanism's incentive guarantee is an empirical quantity: inherited, approximate, and tied to the specific proxy profile that produced it. We give the guarantee a carrier. The three results of this section are proved in Appendix~\ref{app:proofs}.

\begin{definition}[IC certificate]
\label{def:cert}
An \emph{IC certificate} for a wrapped mechanism is a tuple $(\Gamma, h(\pi), \skew{2}\hat\varepsilon, \delta, N)$ asserting: with confidence $1 - \delta$ over $N$ independent type-profile samples from $F$, the estimated maximum manipulation gain satisfies $\max_{i, \theta_i} G_i(\theta_i; \pi) \le \skew{2}\hat\varepsilon$, where $h(\pi)$ is a version hash of the proxy profile.
\end{definition}

The certificate is statistical because the quantity it bounds is not exactly computable at feasible cost, even when the proxy profile is given in full.

\begin{proposition}[Exact certification is intractable]
\label{prop:hardness}
Even with the proxy profile given explicitly, the certified quantity is not polynomial-time computable in the worst case: (i) deciding whether a succinctly represented wrapped mechanism $g \circ \pi$ is exactly dominant-strategy IC is $\mathrm{coNP}$-hard, and (ii) computing the interim manipulation gain $G_i(\theta_i; \pi)$ is $\#\mathrm{P}$-hard. Hence no polynomial-time exact certifier exists unless $\mathrm{P} = \mathrm{NP}$ (resp.\ $\mathrm{FP} = \#\mathrm{P}$).
\end{proposition}

The barrier parallels the counting hardness of Bayesian interim quantities \cite{gnr2015border,conitzersandholm2002complexity}, so the certificate estimates rather than computes. The hardness is in the succinct description: with types and messages given as explicit tables, the manipulation gain is a polynomial-size sum and the check is polynomial. The barrier bites only when the type space is exponential in the input, as it is for a prompt-driven language model (the reductions in Appendix~\ref{app:proofs} make this precise). Approximate incentive compatibility is learnable from sampled types \cite{balcan2019}, and the guarantee transfers directly. The interim gain $G_i(\theta_i; \pi)$ is an expectation under $F(\cdot \mid \theta_i)$, so reading it off unconditioned profile samples assumes a product prior or conditional-sampling access; under a correlated prior one draws the opponents' types conditional on $\theta_i$. Let $d = \mathrm{Pdim}(\mathcal{F})$ be the pseudo-dimension of the class $\mathcal{F}$ of interim manipulation-gain functions induced by the proxy profile and the mechanism, with utilities in $[0, H]$. By Balcan et al.~\cite{balcan2019}, with probability $1 - \delta$ over $N$ independent type-profile samples the empirical maximum manipulation gain estimates the true $\max_{i, \theta_i} G_i(\theta_i; \pi)$ to within an additive
\[
O\!\left(H\sqrt{\tfrac{d}{N}\,\ln\tfrac{N}{d}} \;+\; H\sqrt{\tfrac{1}{N}\,\ln\tfrac{n}{\delta}}\right),
\]
uniformly over the $n$ agents; equivalently, $N = \tilde{O}\big(H^2(d + \ln(n/\delta))/\varepsilon^2\big)$ samples suffice for an $\varepsilon$-accurate certificate. For the bid-scaling proxies of the autobidding model the proxy is a single parameter, the multiplier $\alpha$. Since $\alpha \mapsto G_i(\theta_i; \pi_\alpha)$ is unimodal at each $\theta_i$ (piecewise monotone with $O(1)$ pieces, a $V$ around the loyal multiplier where $G_i = 0$), each level set $\{\alpha : G_i \ge r\}$ is a union of $O(1)$ intervals and the induced family has $d = O(1)$. Then $N = \tilde{O}(H^2/\varepsilon^2)$ samples suffice, the bidder count entering only through the $\ln(n/\delta)$ confidence term. Our contribution is the versioning and drift apparatus below, not the estimation itself.

One caveat needs a calibrated lower bound. Deep-learning regret estimators underestimate true regret \cite{you2026regret}, and the underestimation lives in the per-sample search, which no concentration bound repairs. The remedy is two-layer, each layer one-sided. First, score each sampled profile with a method that does not undershoot the per-profile gain: an exact mixed-integer encoding for piecewise-linear proxies \cite{curry2020certifying}, or a multi-start search seeded to escape poor local optima \cite{you2026regret}. Second, lift the scores to a one-sided population bound with a finite-sample guarantee: the uniform bound above \cite{balcan2019}, a distribution-free conformal upper quantile \cite{angelopoulos2021gentle}, or an empirical-Bernstein bound that tightens when the regret variance is small \cite{maurer2009empirical}. Both layers overstate rather than understate, so the reported $\skew{2}\hat\varepsilon$ can only exceed the true gain. An incentive-compatibility pass then stays sound under an underestimating regret model.

The certificate names a version because the guarantee dies with the version.

\begin{lemma}[Drift]
\label{lem:drift}
Equip each $M_j$ with a metric $d_j$ and suppose interim utilities are $L$-Lipschitz in the message profile: for all $i$, $\theta_i$, and message profiles $m, m'$, $\;|u_i(g(m), \theta_i) - u_i(g(m'), \theta_i)| \le L \sum_j d_j(m_j, m'_j)$. If a proxy update replaces $\pi$ by $\pi'$ with $\max_j \sup_{t} d_j(\pi_j(t), \pi'_j(t)) \le \Delta$, then for every $i$ and $\theta_i$,
\[
G_i(\theta_i; \pi') \;\le\; G_i(\theta_i; \pi) + 2 L n \Delta .
\]
\end{lemma}

\begin{proposition}[Online re-certification under a drift budget]
\label{prop:online}
Consider a stream of proxy versions $\pi^1, \pi^2, \dots$ with per-update action drift $\Delta_t$ (as in Lemma~\ref{lem:drift}) and total drift budget $B_T = \sum_{t < T} \Delta_t$. Fix a slack $s > 0$, advertise the bound $\skew{2}\hat\varepsilon_{\mathrm{last}} + s$ between certificates (the last estimate plus the slack), and re-estimate from $N$ fresh samples whenever $2 L n$ times the drift accumulated since the last certificate would exceed $s$. Then, with $K = \lceil 2 L n\, B_T / s \rceil + 1$ and probability $1 - K\delta$: (i) at every epoch the advertised bound $\skew{2}\hat\varepsilon_{\mathrm{last}} + s$ upper-bounds the true maximum manipulation gain, so the deployed mechanism is never uncertified; and (ii) at most $K$ certificates are issued. Verification cost thus scales with the drift budget $B_T$, not the number of updates: a stream of small updates ($B_T = o(T)$) incurs vanishing amortized cost.
\end{proposition}

In the autobidding instance the drift constant is explicit too: a multiplier change $\Delta\alpha$ moves each bid by at most $\Delta\alpha\,\bar v$ when values are bounded by $\bar v$, so re-certification triggers once the total multiplier drift exceeds $s/(2 L \bar v n)$. Online measurement of IC has been studied for a single fixed auction \cite{feng2019onlineic}. Budgeted re-certification of a versioned mechanism has no classical counterpart, since exact equilibria do not version. A provider's silent model update is an uncertified mechanism change.


\section{Empirical study}
\label{sec:empirics}

The identity predicts that a binding, capability-preserving guardrail leaves $W_i > 0$, so honest reporting is suboptimal and the principal inflates its report to recover the intended action. We test this on production language models used as autobidding proxies. All runs use \iftoggle{anon}{a market-simulation library}{the \texttt{llm-market-sim} library}~\cite{llmmarketsim} and its delegation harness, which sweeps a proxy, analytic or language-model, through a grid of reports and returns the within-range regret and the report inflation; the code and data reproducing every figure and table are archived together~\cite{delegatedplaycode}. The scenario is stylized and the cap is one we impose, not read from any model's safety training, so the study illustrates the mechanism the identity predicts rather than measuring a deployed alignment layer.

\paragraph{Setup.} Each proxy bids for one client in a first-price auction against a single competitor whose bid is uniform on $[0,300]$. The client's value $v$ ranges over $\{100,150,200,250,300\}$, and interim utility is $u(b,v) = (v-b)\min(b/300,1)$, so the unconstrained optimum is $b^\star(v) = v/2$. An alignment-style soft cap compresses the proxy's bid, $c(b) = b$ for $b \le 40$ and $c(b) = 40 + 0.4\,(b-40)$ above, leaving the optimum reachable through a higher report, up to the models' own bid ceiling. The model is instructed only to maximize the client's expected surplus\footnote{The system prompt reads, in part: ``Bid to maximize the client's expected surplus, the value minus the price paid weighted by the chance of winning.''} and is not shown the cap; it is applied to the bid the model returns. For each model we sweep the reported value over $25$ evenly spaced reports ($50, 75, \dots, 650$), query the model once per report, parse its bid (the first number in the reply, clamped to $[0, 300]$; an unparseable reply scores as a zero bid, though none occurred, $0$ of $625$ calls), and estimate $W_i$ as the gap between the best reachable payoff on the grid and the honest-report ($t = v$) payoff. Figure~\ref{fig:vignette} sketches this report-to-bid mapping at $v = 200$. Decoding is at temperature $0$ for the standard chat models; the reasoning models (GPT-5 series, o-series) do not accept a temperature and run at their fixed default. Each report is a single draw; the nonzero-temperature reasoning models nonetheless returned the rational bid at all twenty-five reports, and re-querying five of them at five reports five times each returned identical bids every time (zero spread), so these decoders are effectively deterministic here and the single draw is representative. Every request's decode parameters and both prompts are recorded in the reproducibility archive. Formally the type space is this report grid with the prior concentrated on the five values, so every swept report is a valid type-report. The grid is wide enough that the first-price argmax is interior at every value, and at $v = 300$ it is the report ($600$) at which the bid reaches its ceiling of $300$, so the frontier there is set by the models' own bid ceiling rather than the grid boundary.

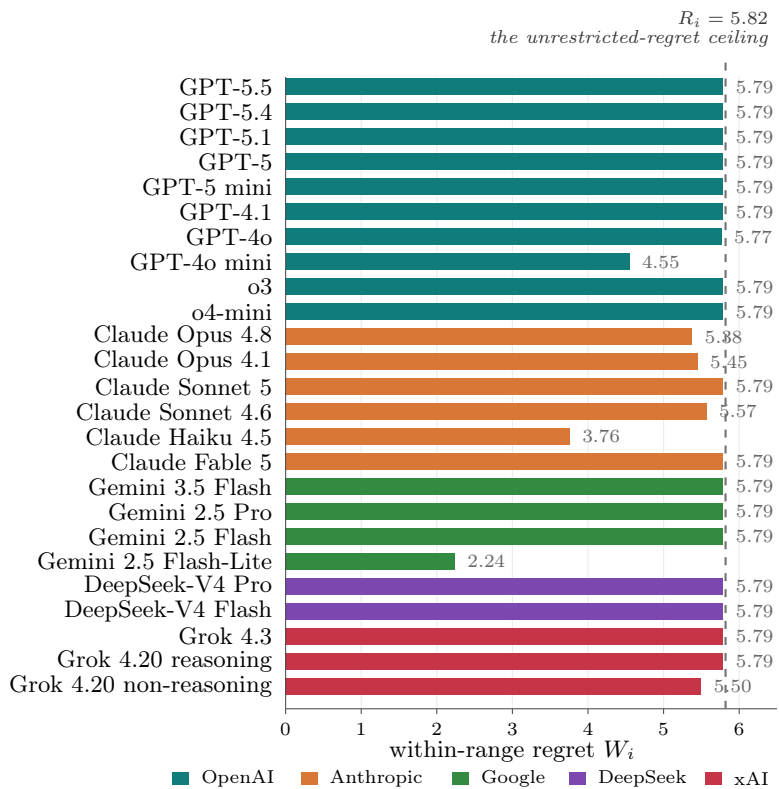
\begin{figure}[tbp]
\centering
\begin{tikzpicture}[font=\footnotesize]
\draw[black!8] (1.00,0.03) -- (1.00,8.38);
\draw[black!8] (2.00,0.03) -- (2.00,8.38);
\draw[black!8] (3.00,0.03) -- (3.00,8.38);
\draw[black!8] (4.00,0.03) -- (4.00,8.38);
\draw[black!8] (5.00,0.03) -- (5.00,8.38);
\draw[black!8] (6.00,0.03) -- (6.00,8.38);
\fill[oai] (0,8.14) rectangle (5.79,8.36);
\node[left,font=\footnotesize] at (-0.06,8.25) {GPT-5.5};
\node[right,font=\scriptsize,black!60] at (5.84,8.25) {5.79};
\fill[oai] (0,7.81) rectangle (5.79,8.03);
\node[left,font=\footnotesize] at (-0.06,7.92) {GPT-5.4};
\node[right,font=\scriptsize,black!60] at (5.84,7.92) {5.79};
\fill[oai] (0,7.48) rectangle (5.79,7.70);
\node[left,font=\footnotesize] at (-0.06,7.59) {GPT-5.1};
\node[right,font=\scriptsize,black!60] at (5.84,7.59) {5.79};
\fill[oai] (0,7.15) rectangle (5.79,7.37);
\node[left,font=\footnotesize] at (-0.06,7.26) {GPT-5};
\node[right,font=\scriptsize,black!60] at (5.84,7.26) {5.79};
\fill[oai] (0,6.82) rectangle (5.79,7.04);
\node[left,font=\footnotesize] at (-0.06,6.93) {GPT-5 mini};
\node[right,font=\scriptsize,black!60] at (5.84,6.93) {5.79};
\fill[oai] (0,6.49) rectangle (5.79,6.71);
\node[left,font=\footnotesize] at (-0.06,6.60) {GPT-4.1};
\node[right,font=\scriptsize,black!60] at (5.84,6.60) {5.79};
\fill[oai] (0,6.16) rectangle (5.77,6.38);
\node[left,font=\footnotesize] at (-0.06,6.27) {GPT-4o};
\node[right,font=\scriptsize,black!60] at (5.82,6.27) {5.77};
\fill[oai] (0,5.83) rectangle (4.55,6.05);
\node[left,font=\footnotesize] at (-0.06,5.94) {GPT-4o mini};
\node[right,font=\scriptsize,black!60] at (4.60,5.94) {4.55};
\fill[oai] (0,5.50) rectangle (5.79,5.72);
\node[left,font=\footnotesize] at (-0.06,5.61) {o3};
\node[right,font=\scriptsize,black!60] at (5.84,5.61) {5.79};
\fill[oai] (0,5.17) rectangle (5.79,5.39);
\node[left,font=\footnotesize] at (-0.06,5.28) {o4-mini};
\node[right,font=\scriptsize,black!60] at (5.84,5.28) {5.79};
\fill[ant] (0,4.84) rectangle (5.38,5.06);
\node[left,font=\footnotesize] at (-0.06,4.95) {Claude Opus 4.8};
\node[right,font=\scriptsize,black!60] at (5.43,4.95) {5.38};
\fill[ant] (0,4.51) rectangle (5.45,4.73);
\node[left,font=\footnotesize] at (-0.06,4.62) {Claude Opus 4.1};
\node[right,font=\scriptsize,black!60] at (5.50,4.62) {5.45};
\fill[ant] (0,4.18) rectangle (5.79,4.40);
\node[left,font=\footnotesize] at (-0.06,4.29) {Claude Sonnet 5};
\node[right,font=\scriptsize,black!60] at (5.84,4.29) {5.79};
\fill[ant] (0,3.85) rectangle (5.57,4.07);
\node[left,font=\footnotesize] at (-0.06,3.96) {Claude Sonnet 4.6};
\node[right,font=\scriptsize,black!60] at (5.62,3.96) {5.57};
\fill[ant] (0,3.52) rectangle (3.76,3.74);
\node[left,font=\footnotesize] at (-0.06,3.63) {Claude Haiku 4.5};
\node[right,font=\scriptsize,black!60] at (3.81,3.63) {3.76};
\fill[ant] (0,3.19) rectangle (5.79,3.41);
\node[left,font=\footnotesize] at (-0.06,3.30) {Claude Fable 5};
\node[right,font=\scriptsize,black!60] at (5.84,3.30) {5.79};
\fill[goo] (0,2.86) rectangle (5.79,3.08);
\node[left,font=\footnotesize] at (-0.06,2.97) {Gemini 3.5 Flash};
\node[right,font=\scriptsize,black!60] at (5.84,2.97) {5.79};
\fill[goo] (0,2.53) rectangle (5.79,2.75);
\node[left,font=\footnotesize] at (-0.06,2.64) {Gemini 2.5 Pro};
\node[right,font=\scriptsize,black!60] at (5.84,2.64) {5.79};
\fill[goo] (0,2.20) rectangle (5.79,2.42);
\node[left,font=\footnotesize] at (-0.06,2.31) {Gemini 2.5 Flash};
\node[right,font=\scriptsize,black!60] at (5.84,2.31) {5.79};
\fill[goo] (0,1.87) rectangle (2.24,2.09);
\node[left,font=\footnotesize] at (-0.06,1.98) {Gemini 2.5 Flash-Lite};
\node[right,font=\scriptsize,black!60] at (2.29,1.98) {2.24};
\fill[dee] (0,1.54) rectangle (5.79,1.76);
\node[left,font=\footnotesize] at (-0.06,1.65) {DeepSeek-V4 Pro};
\node[right,font=\scriptsize,black!60] at (5.84,1.65) {5.79};
\fill[dee] (0,1.21) rectangle (5.79,1.43);
\node[left,font=\footnotesize] at (-0.06,1.32) {DeepSeek-V4 Flash};
\node[right,font=\scriptsize,black!60] at (5.84,1.32) {5.79};
\fill[xai] (0,0.88) rectangle (5.79,1.10);
\node[left,font=\footnotesize] at (-0.06,0.99) {Grok 4.3};
\node[right,font=\scriptsize,black!60] at (5.84,0.99) {5.79};
\fill[xai] (0,0.55) rectangle (5.79,0.77);
\node[left,font=\footnotesize] at (-0.06,0.66) {Grok 4.20 reasoning};
\node[right,font=\scriptsize,black!60] at (5.84,0.66) {5.79};
\fill[xai] (0,0.22) rectangle (5.50,0.44);
\node[left,font=\footnotesize] at (-0.06,0.33) {Grok 4.20 non-reasoning};
\node[right,font=\scriptsize,black!60] at (5.55,0.33) {5.50};
\draw[black!70] (0,0.03) -- (6.50,0.03);
\draw[black!70] (0,0.03) -- (0,8.38);
\draw[black!70] (0.00,0.03)--(0.00,-0.04) node[below,font=\scriptsize,black]{0};
\draw[black!70] (1.00,0.03)--(1.00,-0.04) node[below,font=\scriptsize,black]{1};
\draw[black!70] (2.00,0.03)--(2.00,-0.04) node[below,font=\scriptsize,black]{2};
\draw[black!70] (3.00,0.03)--(3.00,-0.04) node[below,font=\scriptsize,black]{3};
\draw[black!70] (4.00,0.03)--(4.00,-0.04) node[below,font=\scriptsize,black]{4};
\draw[black!70] (5.00,0.03)--(5.00,-0.04) node[below,font=\scriptsize,black]{5};
\draw[black!70] (6.00,0.03)--(6.00,-0.04) node[below,font=\scriptsize,black]{6};
\node[below,font=\footnotesize] at (3.00,-0.27) {within-range regret $W_i$};
\draw[black!55,dashed,thick] (5.82,0.03) -- (5.82,8.65);
\node[anchor=south east,align=right,font=\scriptsize,text=black!78] at (6.50,8.67) {$R_i = 5.82$\\[-1pt]{\itshape the unrestricted-regret ceiling}};
\fill[oai] (-1.50,-0.97) rectangle (-1.28,-0.81);
\node[right,font=\scriptsize] at (-1.23,-0.89) {OpenAI};
\fill[ant] (0.20,-0.97) rectangle (0.42,-0.81);
\node[right,font=\scriptsize] at (0.47,-0.89) {Anthropic};
\fill[goo] (2.20,-0.97) rectangle (2.42,-0.81);
\node[right,font=\scriptsize] at (2.47,-0.89) {Google};
\fill[dee] (3.75,-0.97) rectangle (3.97,-0.81);
\node[right,font=\scriptsize] at (4.02,-0.89) {DeepSeek};
\fill[xai] (5.55,-0.97) rectangle (5.77,-0.81);
\node[right,font=\scriptsize] at (5.82,-0.89) {xAI};
\end{tikzpicture}
\caption{Within-range regret $W_i$ for 25 production language models across five providers, each an autobidding proxy under the alignment-style soft cap (mean over true values $v \in \{100, \dots, 300\}$; exact values labeled). Honest reporting is suboptimal on every model: the capable ones bid the rational $v/2$ and cluster at $W_i = 5.79$, just under the analytic optimal-play regret $R_i = 5.82$ (dashed; a mean over $v$, where the $R_i$ of Section~\ref{sec:model} is a supremum over types, and a ceiling for rational play, not for arbitrary proxies, GPT-4o's menu-arm $W_i = 9.39$ already exceeding it), while the smaller models bid less precisely and sit lower. Report inflation, discussed in the text, runs from $1.2\times$ to $1.8\times$; per-model $W_i$ and inflation are in Appendix Table~\ref{tab:sweep-detail}.}
\label{fig:sweep}
\end{figure}

\begin{figure}[tbp]
\centering
\begin{tikzpicture}
\begin{axis}[
  width=0.62\linewidth, height=4.3cm,
  xlabel={report $t$}, ylabel={bid the mechanism sees},
  xmin=50, xmax=500, ymin=0, ymax=126,
  xtick={200,380}, xticklabels={$v$,$t^\star$},
  ytick={64,100}, yticklabels={$64$,$v/2$},
  tick label style={font=\footnotesize}, label style={font=\footnotesize},
  axis lines=left, clip=false,
]
\addplot[red!70!black, dashed, thick, domain=50:500, samples=2] {100};
\addplot[blue!70!black, very thick, domain=50:80, samples=2] {x/2};
\addplot[blue!70!black, very thick, domain=80:500, samples=2] {24+0.2*x};
\addplot[only marks, mark=*, mark size=1.7pt, black] coordinates {(200,64) (380,100)};
\draw[densely dotted] (axis cs:200,0) -- (axis cs:200,64);
\draw[densely dotted] (axis cs:380,0) -- (axis cs:380,100);
\end{axis}
\end{tikzpicture}
\caption{The mechanism behind Figure~\ref{fig:sweep}, at true value $v = 200$: the guardrailed bid the mechanism sees against the reported value $t$ is the soft cap applied to the rational bid $t/2$. The honest report $t = v$ yields a bid below the optimum $v/2$, and an inflated report $t^\star$ recovers it.}
\label{fig:vignette}
\end{figure}
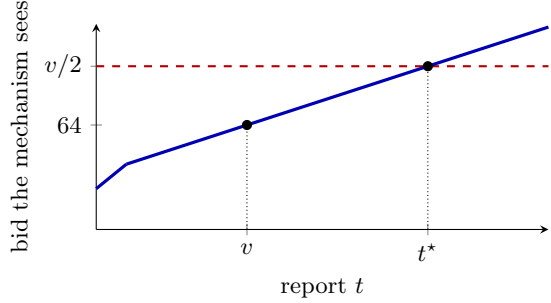

\paragraph{The incentive appears on every model (Figure~\ref{fig:sweep}).}
Measured on 25 production language models across five providers under the alignment-style soft cap, the binding, capability-preserving corner gives $W_i > 0$ with report inflation above one on every model, and the cap binds at every value on all but one (Gemini 2.5 Flash-Lite, loose enough that honest reporting is already optimal at $v = 100$). The disloyalty is the composite's, not the base model's: instructed to maximize client surplus, the capable models play the rational bid, and the cap applied to it displaces the optimum, so the wrapped proxy $c \circ \pi$ has $W_i > 0$ on every model. Seventeen produce the rational bid capped at the competitor's support maximum, effective bid $\min(\lfloor t/2 \rceil, 300)$ ($\lfloor \cdot \rceil$ rounds to the nearest integer), at all twenty-five reports, and so share $W_i = 5.79$ and inflation $1.76\times$ exactly. Bidding above $300$ is dominated in first price; the models emit $300$ directly at the high reports, and the parser clamps the rare overshoot to the same ceiling, so at $v = 300$ the frontier is $c(300) = 144$ rather than the $c(325) = 154$ that untruncated play would reach. That shared value is the within-range regret of this bid function under the cap, a check on the harness; the untruncated $t/2$ would give $5.82$. The unrestricted regret of the same play is $R_i = 5.82$, so $W_i$ nearly saturates $R_i$. The report channel can inflate to approach the unconstrained optimum, and only the cap keeps it short. Within-range regret ranges from $2.2$ to $5.8$; the smaller models (GPT-4o mini, Claude Haiku 4.5, Gemini 2.5 Flash-Lite) show lower $W_i$ because they bid imprecisely, over- or under-shooting $v/2$, not because honest reporting is nearer optimal for them. GPT-4o sits just below the cluster at $5.77$, its bid at $v = 250$ a shade off the rational $v/2$. The identical rows span distinct models and providers, so they reflect the same rational strategy rather than coincidence. Every call records its decode parameters, system and user prompts, and raw response, so each number is reproducible.

\paragraph{The incentive persists across environments (Table~\ref{tab:robust}).}
The soft cap is one environment; the incentive it induces is not specific to it. On a representative model per provider we re-run the sweep under a narrower competitor distribution and under a second-price payment rule, where honest bidding is optimal absent the cap. That second rule is the decisive test: it removes every misreporting incentive except the guardrail's own, and the cap still forces report inflation on every model. In both the cap still binds and report inflation persists, with within-range regret positive on every model: mean inflation $1.55\times$ under the narrower competitor range and $1.54\times$ under second-price, against $1.74\times$ for these five models in the headline first-price setup. Within-range regret is reported in payoff units, which scale with the environment, so inflation is the comparable measure across rows; it barely moves. The drop under the narrower range is the same truncation seen in the headline sweep: bids cap at the support maximum $200$, so for $v \in \{250, 300\}$ the frontier is $c(200) = 104$ and less report inflation reaches it. In that decisive case, with payoff $u(b,v) = \min(b/300,1)\,(v - b/2)$ for $b \le 300$, all five models bid value-faithfully ($b = \min(t,300)$), so for $v \ge 150$ the reachable frontier is the capped top bid $c(300) = 144$, while at $v = 100$ it is $c(200) = 104$, the reachable bid nearest the value. The within-range regret is measured to the per-value frontier, mean $5.44$. The regret to the uncapped optimum $b = v$ is $18.4$. $W_i$ correctly excludes it because that bid is not reachable through any report.

\begin{table}[tbp]
\centering
\small
\caption{Robustness of the report-inflation incentive across environments, mean over one representative model per provider (GPT-4o, Claude Sonnet 5, Gemini 2.5 Flash, DeepSeek-V4 Flash, Grok 4.3). The incentive persists in every environment; inflation, the scale-free measure, is nearly unchanged. ``$W_i > 0$'' is the fraction of values at which honest reporting is suboptimal, the failure of Truthfulness, not merely where the cap compresses the bid, which happens at every value. It is distinct from \emph{Binding} in Section~\ref{sec:limits} (displacement of the honest action from the unguarded optimum $a^\star$): the second-price arm at $v = 300$ reaches the capped frontier ($W_i = 0$) yet leaves $a^\star = 300$ displaced, so that value is capability-losing, not non-binding.}
\label{tab:robust}
\begin{tabular*}{\textwidth}{@{\extracolsep{\fill}} l c c c}
\toprule
Environment & mean $W_i$ & mean inflation & $W_i > 0$ \\
\midrule
First-price, competitor $\sim U[0,300]$ (headline) & $5.79$ & $1.74\times$ & $100\%$ \\
First-price, competitor $\sim U[0,200]$ & $6.17$ & $1.55\times$ & $100\%$ \\
Second-price, competitor $\sim U[0,300]$ & $5.44$ & $1.54\times$ & $80\%$ \\
\bottomrule
\end{tabular*}
\end{table}

\paragraph{Placement decides the corner (Table~\ref{tab:ablation}).}
Proposition~\ref{prop:placement} predicts that the \emph{same} cap lands in a different corner by placement: output-clipping, the cap applied to the proxy's chosen bid, leaves $W_i > 0$, while constraint-aware decoding, the proxy selecting the best \emph{allowed} bid, restores $W_i = 0$. We run both on the same models. Output-clipping is the headline sweep; for the constraint-aware arm we present each model the menu $\{t/4, t/2, 3t/4, t\}$ of allowed bids built from the reported value $t$, which at the honest report $t = v$ offers the interim optimum $v/2$ (items rounded to integers), and ask which maximizes client surplus, exactly Proposition~\ref{prop:placement}'s "selects the best allowed output"; the reply is snapped to the nearest menu item and an unparseable reply to the lowest (none needed the fallback), and $W_i$ is the shortfall of the honest-report choice from the best allowed bid, which upper-bounds the within-range regret and equals it here, since $v/2$ is recovered by a report. This carries no cap arithmetic to misread, unlike telling the model a formula. The two arms differ in allowed set as well as placement, the cap's compressed range against the four-point menu; $v/2$ lies in both at every value below $300$, so the comparison isolates placement, and at $v = 300$ the clip frontier $c(300) = 144$ leaves a $0.12$ shortfall ($0.16\%$), Proposition~\ref{prop:quant-trilemma}'s interior rather than a pure corner. Constraint-aware placement drops within-range regret to zero on four of the five models, which select $v/2$ at the honest report, against $W_i \approx 5.79$ under output-clipping, and the mean falls from $5.79$ to $1.88$. GPT-4o is the exception: it over-selects, choosing $3v/4$ rather than the optimal $v/2$, so $W_i = u(v/2, v) - u(3v/4, v) = 9.39$ (menu integers rounded half-to-even); the report that recovers its optimum is a \emph{deflation} ($t^\star \approx 2v/3$, where its $3t/4$ choice lands on $v/2$), the opposite sign to the clipping arm's inflation, and placement alone does not rescue a proxy that misoptimizes. With the optimum reachable in both arms, where the constraint sits decides whether truthfulness survives. Randomizing the menu order rules out option-position bias: across shuffled orders the four loyal models select $v/2$ in all $15$ trials, and GPT-4o still over-selects $3v/4$ in $14$ of $15$, its lone $v/2$ pick at $v = 300$ ($W_i = 8.14$ against the ascending $9.39$), so neither the restored truthfulness nor GPT-4o's anomaly is an artifact of where the optimum sits in the list.

\begin{table}[tbp]
\centering
\small
\caption{Placement ablation (Proposition~\ref{prop:placement}): within-range regret under output-clipping (the cap applied to the model's bid) and constraint-aware decoding (the model selects the best allowed bid), for the same five models. Placement sets the corner: constraint-aware restores $W_i = 0$ on four of five; GPT-4o over-selects even from the allowed menu.}
\label{tab:ablation}
\begin{tabular*}{\textwidth}{@{\extracolsep{\fill}} l c c}
\toprule
Model & output-clipping $W_i$ & constraint-aware $W_i$ \\
\midrule
GPT-4o & $5.77$ & $9.39$ \\
Claude Sonnet 5 & $5.79$ & $0.00$ \\
Gemini 2.5 Flash & $5.79$ & $0.00$ \\
DeepSeek-V4 Flash & $5.79$ & $0.00$ \\
Grok 4.3 & $5.79$ & $0.00$ \\
\midrule
mean & $5.79$ & $1.88$ \\
\bottomrule
\end{tabular*}
\end{table}

\paragraph{Disclosing the cap reaches the corner only when the arithmetic does.}
Constraint-aware decoding can also be prompted directly: a third arm discloses the cap \emph{formula} and asks the model to bid accounting for it, reaching the truthful corner by re-optimization rather than menu selection. On the same five models the result splits on arithmetic. Gemini 2.5 Flash and Claude Sonnet 5 invert the cap and collapse to $W_i = 0.00$ and $0.09$ (inflation $1.00\times$ and $1.25\times$), reaching the corner through the prompt alone; the other three mis-invert and \emph{overcorrect} past their output-clipping value, to $W_i = 14.1$ (DeepSeek-V4 Flash), $9.50$ (GPT-4o), and $8.98$ (Grok 4.3). The menu of Table~\ref{tab:ablation} removes this arithmetic, which is why it restores $W_i = 0$ on four models rather than roughly two: it rescues DeepSeek-V4 Flash and Grok 4.3, which can rank the allowed bids but not invert the cap. Naming the constraint and trusting the model to re-optimize reaches the corner only when the model computes the inverse; offering the allowed set reaches it whenever the model can rank. An intermediate arm, disclosing the cap's existence but not its formula, would separate awareness of the constraint from the ability to invert it; we leave it to future work.

\paragraph{The third corner: a hard cap forfeits capability, not truthfulness.}
The soft cap keeps the optimum reachable, bar the $0.16\%$ ceiling shortfall at $v = 300$, and so occupies the capability-preserving corner. A \emph{hard} cap $c(b) = \min(b, 40)$ that removes the optimum from reach occupies the opposite one. Re-scoring the same collected bids under the hard cap, all 25 models have $W_i = 0$: honest reporting already reaches the capped frontier, so nothing is gained by inflating. The mean capability loss is $16.2$ payoff units, the surplus the cap destroys by putting $v/2$ out of reach. Truthful and capability-losing, the hard cap is the capability-removing output-clip of Proposition~\ref{prop:quant-trilemma}, spending its whole loss budget on capability rather than truthfulness. All three corners of Figure~\ref{fig:trilemma}(b) now appear empirically.

Three further checks are in Appendix~\ref{app:empirics}: the $R_i - W_i$ gap across $60{,}000$ random games (Fig.~\ref{fig:rwgap}), the analytic soft-cap inflation profile (Fig.~\ref{fig:inflation}), and the sampling certificate's convergence and re-certification under drift (Fig.~\ref{fig:certification}).


\section{Discussion}
\label{sec:discussion}

\paragraph{What we found.}
Delegation leaves a principal one lever, the report it hands its proxy. A single quantity decides whether that lever keeps it honest: the proxy's within-range regret. Honest self-description is optimal exactly when the proxy already plays the best action it can reach (Theorem~\ref{thm:ttp}), the incentive deployed elicitation proxies \cite{huang2025elicitation} leave unmodeled. The identity subsumes the autobidding incentive-compatibility results and turns guardrails into mechanism-relevant objects: none binds, stays truthful, and preserves capability at once (Theorem~\ref{thm:trilemma}). Which one a design forfeits depends on its placement and its allowed set (Proposition~\ref{prop:placement}). Honest self-description thus stops being an assumption and becomes a property a platform measures: \#\textrm{P}-hard to compute, but estimable from samples and maintainable as the proxy drifts. Run on production language models from five providers under an alignment-style cap we impose, that estimate recovers the predicted incentive on every model (Figure~\ref{fig:sweep}).

\paragraph{Loyalty and what to reveal.}
Every deployment takes the proxy to act for its principal; none certifies that it does. The results need less, holding against whatever proxy a principal faces. A disloyal proxy does not break them. It becomes the object they describe, and within-range regret measures exactly its misalignment (Lemma~\ref{lem:loyalty}). Certifying \emph{which} objective a proxy optimizes is the framework's most consequential open problem, and it governs a second design choice, the report language. That language fixes $\Reach_i = \pi_i(\Theta_i)$, so a principal should reveal exactly enough that its optimum is reachable ($a^\star \in \Reach_i$) and no more. Reach beyond the optimum only enlarges the surface a guardrail can bind, and against a proxy one cannot certify loyal, it converts into within-range regret. The worth of a loyalty certificate is exactly the reach it makes safe to share. Choosing how much to reveal is the classical delegation-set problem \cite{holmstrom1984,alonsomatouschek2008,armstrong2010delegated,kleinberg2018delegated} from the principal's side, an instance of the aligned delegation of Frankel \cite{frankel2014aligned} with within-range regret playing the role of the agent's bias. The wrap also inherits the principle's classical limits, limited commitment \cite{bester2001,dovalskreta2022} and collusion among proxies \cite{fish2024collusion}.

\paragraph{Open problems.}
\emph{Measured manipulation}: measure $W_i$ from a model's own refusal behavior rather than an imposed cap, the constraint-aware and output-clipping contrast of Proposition~\ref{prop:placement} now shown only under a cap we impose. \emph{Equilibrium and welfare}: by Lemma~\ref{lem:coarsen} an input-game equilibrium corresponds to one of the coarsened base game. Whether the inflated profile in which every principal has $W_i > 0$ is itself an equilibrium, and its revenue and welfare against the honest benchmark, remain open. \emph{Beyond auctions}: instantiate the identity in matching and bargaining, the real test of its generality. \emph{Certified loyalty}: certify that a proxy serves its principal rather than the platform, the assumption every deployment makes and none checks. Once loyalty and manipulation are both certifiable, the trilemma becomes a navigable constraint, and the design of proxy ecosystems, who certifies whom and at what price, becomes mechanism design in its own right.

\iftoggle{draftnotes}{%
\par\medskip
{\color{red}%
\noindent\textbf{[draft note: empirical validation program; strip for submission]}
\begin{itemize}\setlength{\itemsep}{1pt}
  \item \textbf{Done.} Truth-to-proxy identity and inheritance (Thm.~\ref{thm:ttp}, Prop.~\ref{prop:inherit}): $G_i = W_i$ and $G_i \le R_i$ over $60{,}000$ random finite games and the worked first-price example, $0$ violations (\texttt{validate\_truth\_to\_proxy.py}).
  \item \textbf{The trilemma (Thm.~\ref{thm:trilemma}).} Numerical illustration of the three corners on the bid-cap example: a binding, capability-preserving guardrail is forced to $W_i > 0$, and the truthful corner forfeits capability.
  \item \textbf{Guardrails (Cor.~\ref{cor:guardrails}).} Bid-cap simulation: input inflation concentrated on principals where the cap binds.
  \item \textbf{Realized proxies.} No-regret learners as proxies; measure $W_i$ from their play and confirm the wrapped mechanism's measured IC slack tracks the measured $\varepsilon$ (empirical Prop.~\ref{prop:inherit}).
  \item \textbf{Certification (Props.~\ref{prop:hardness},~\ref{prop:online}).} Sampling certificate tracks the true $\max_i G_i$; re-certification count scales with cumulative drift $B_T$.
  \item \textbf{Generality (top reviewer risk).} Non-auction instantiation (matching or contract) re-deriving $G_i = W_i$.
  \item \textbf{Stretch.} LLM-proxy demonstration with empirically measured within-range regret, tying the framework to the deployed motivation.
\end{itemize}%
}%
\par\medskip
}{}


\bibliographystyle{splncs04}
\bibliography{bibliography}

\clearpage
\appendix
\section{Additional empirical detail}
\label{app:empirics}

This appendix collects the supporting figures for the empirical study of Section~\ref{sec:empirics}: the within-range gap across random games, the analytic soft-cap profile, and the sampling certificate's behaviour. All runs use \iftoggle{anon}{a market-simulation library}{the \texttt{llm-market-sim} library}~\cite{llmmarketsim} and its delegation harness, archived with the reproduction code~\cite{delegatedplaycode}. Figure~\ref{fig:sweep} plots the per-model within-range regret; the exact values and the report inflation are collected in Table~\ref{tab:sweep-detail}. The language-model sweep ran between 11 and 14 July 2026; each call's model identifier, a provider alias for most models and a dated snapshot for others, and its UTC timestamp are recorded in the raw log, so a provider's silent update to any alias is, in the terms of Definition~\ref{def:cert}, an uncertified change to $h(\pi)$.

\begin{table}[!ht]
\centering
\small
\caption{Per-model within-range regret $W_i$ and report inflation for the 25 models of Figure~\ref{fig:sweep} (mean over true values $v \in \{100, \dots, 300\}$).}
\label{tab:sweep-detail}
\begin{tabular*}{\textwidth}{@{\extracolsep{\fill}} l l r r}
\toprule
Model & Provider & $W_i$ & Inflation \\
\midrule
GPT-5.5 & OpenAI & $5.79$ & $1.76\times$ \\
GPT-5.4 & OpenAI & $5.79$ & $1.76\times$ \\
GPT-5.1 & OpenAI & $5.79$ & $1.76\times$ \\
GPT-5 & OpenAI & $5.79$ & $1.76\times$ \\
GPT-5 mini & OpenAI & $5.79$ & $1.76\times$ \\
GPT-4.1 & OpenAI & $5.79$ & $1.76\times$ \\
GPT-4o & OpenAI & $5.77$ & $1.67\times$ \\
GPT-4o mini & OpenAI & $4.55$ & $1.22\times$ \\
o3 & OpenAI & $5.79$ & $1.76\times$ \\
o4-mini & OpenAI & $5.79$ & $1.76\times$ \\
\midrule
Claude Opus 4.8 & Anthropic & $5.38$ & $1.51\times$ \\
Claude Opus 4.1 & Anthropic & $5.45$ & $1.54\times$ \\
Claude Sonnet 5 & Anthropic & $5.79$ & $1.76\times$ \\
Claude Sonnet 4.6 & Anthropic & $5.57$ & $1.44\times$ \\
Claude Haiku 4.5 & Anthropic & $3.76$ & $1.35\times$ \\
Claude Fable 5 & Anthropic & $5.79$ & $1.76\times$ \\
\midrule
Gemini 3.5 Flash & Google & $5.79$ & $1.76\times$ \\
Gemini 2.5 Pro & Google & $5.79$ & $1.76\times$ \\
Gemini 2.5 Flash & Google & $5.79$ & $1.76\times$ \\
Gemini 2.5 Flash-Lite & Google & $2.24$ & $1.38\times$ \\
\midrule
DeepSeek-V4 Pro & DeepSeek & $5.79$ & $1.76\times$ \\
DeepSeek-V4 Flash & DeepSeek & $5.79$ & $1.76\times$ \\
\midrule
Grok 4.3 & xAI & $5.79$ & $1.76\times$ \\
Grok 4.20 (reasoning) & xAI & $5.79$ & $1.76\times$ \\
Grok 4.20 (non-reasoning) & xAI & $5.50$ & $1.73\times$ \\
\bottomrule
\end{tabular*}
\end{table}

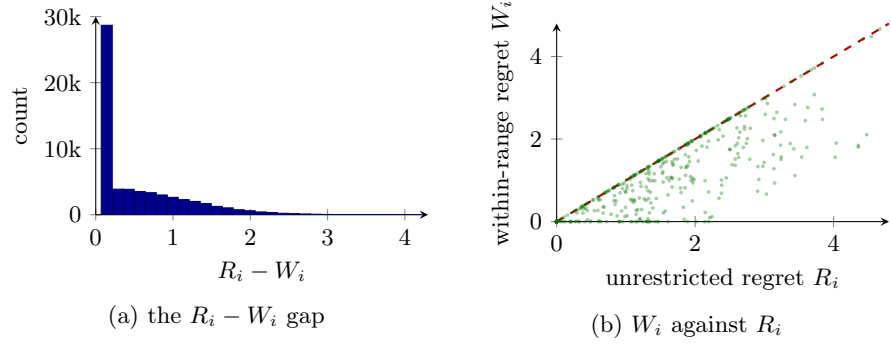
\begin{figure}[tbp]
\centering
\begin{minipage}{0.49\textwidth}\centering
\begin{tikzpicture}
\begin{axis}[width=\linewidth, height=4.2cm,
  xlabel={$R_i - W_i$}, ylabel={count},
  xmin=0, xmax=4.3, ymin=0, ymax=30000,
  ytick={0,10000,20000,30000}, yticklabels={$0$,$10$k,$20$k,$30$k}, scaled y ticks=false,
  tick label style={font=\footnotesize}, label style={font=\footnotesize}, axis lines=left]
\addplot[ybar interval, fill=blue!55!black, draw=blue!40!black] table[x=x, y=count] {figures/rw_gap_hist.dat};
\end{axis}
\end{tikzpicture}
\\[2pt] {\footnotesize (a) the $R_i - W_i$ gap}
\end{minipage}\hfill
\begin{minipage}{0.49\textwidth}\centering
\begin{tikzpicture}
\begin{axis}[width=\linewidth, height=4.2cm,
  xlabel={unrestricted regret $R_i$}, ylabel={within-range regret $W_i$},
  xmin=0, xmax=4.8, ymin=0, ymax=4.8, xtick={0,2,4}, ytick={0,2,4},
  tick label style={font=\footnotesize}, label style={font=\footnotesize}, axis lines=left]
\addplot[only marks, mark=*, mark size=0.5pt, color=green!45!black, opacity=0.35] table[x=R, y=W] {figures/rw_gap_scatter.dat};
\addplot[red!70!black, dashed, thick, domain=0:4.8, samples=2] {x};
\end{axis}
\end{tikzpicture}
\\[2pt] {\footnotesize (b) $W_i$ against $R_i$}
\end{minipage}
\caption{(a) The $R_i - W_i$ gap across $60{,}000$ random games and (b) $W_i$ against $R_i$ with the $W = R$ diagonal. The proxy's range removes real regret in most games, so $W_i$, not $R_i$, governs misreporting.}
\label{fig:rwgap}
\end{figure}

\paragraph{Within-range regret is genuinely smaller than unrestricted regret (Fig.~\ref{fig:rwgap}).}
Across $60{,}000$ random finite Bayesian games with random proxy ranges, $W_i$ is strictly below $R_i$ in $58\%$ of instances, with $W_i = 0$ outright in $20\%$ (the reachable set already contains the honest optimum), mean $R_i = 1.64$ against mean $W_i = 1.17$. The identity $G_i = W_i$ and the bound $G_i \le R_i$ hold with no violations, computed through independent code paths.

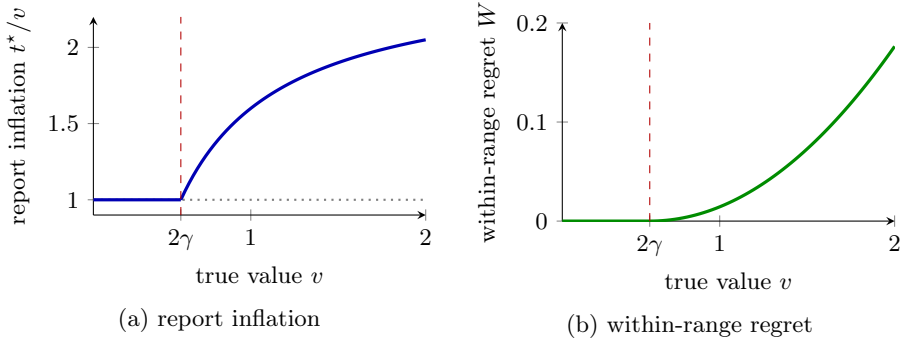
\begin{figure}[tbp]
\centering
\begin{minipage}{0.49\textwidth}\centering
\begin{tikzpicture}
\begin{axis}[width=\linewidth, height=4.2cm,
  xlabel={true value $v$}, ylabel={report inflation $t^\star/v$},
  xmin=0.1, xmax=2, ymin=0.9, ymax=2.2,
  xtick={0.6,1,2}, xticklabels={$2\gamma$,$1$,$2$}, ytick={1,1.5,2},
  tick label style={font=\footnotesize}, label style={font=\footnotesize}, axis lines=left]
\addplot[gray, dotted, thick, domain=0.1:2, samples=2] {1};
\addplot[blue!70!black, very thick, domain=0.1:0.6, samples=2] {1};
\addplot[blue!70!black, very thick, domain=0.6:2, samples=60] {2.5-0.9/x};
\draw[red!70!black, dashed] (axis cs:0.6,0.9) -- (axis cs:0.6,2.2);
\end{axis}
\end{tikzpicture}
\\[2pt] {\footnotesize (a) report inflation}
\end{minipage}\hfill
\begin{minipage}{0.49\textwidth}\centering
\begin{tikzpicture}
\begin{axis}[width=\linewidth, height=4.2cm,
  xlabel={true value $v$}, ylabel={within-range regret $W$},
  xmin=0.1, xmax=2, ymin=0, ymax=0.2,
  xtick={0.6,1,2}, xticklabels={$2\gamma$,$1$,$2$}, ytick={0,0.1,0.2},
  tick label style={font=\footnotesize}, label style={font=\footnotesize}, axis lines=left]
\addplot[green!55!black, very thick, domain=0.1:0.6, samples=2] {0};
\addplot[green!55!black, very thick, domain=0.6:2, samples=60] {0.09*(x-0.6)^2};
\draw[red!70!black, dashed] (axis cs:0.6,0) -- (axis cs:0.6,0.2);
\end{axis}
\end{tikzpicture}
\\[2pt] {\footnotesize (b) within-range regret}
\end{minipage}
\caption{(a) Report inflation and (b) within-range regret against the true value $v$ for the soft-cap proxy, both flat until the cap binds at $v = 2\gamma$ (dashed) and rising past it; within-range regret grows as $W = 0.09\,(v - 2\gamma)^2$ in this example, the quadratic being the $(\varepsilon, \tfrac{\mu}{2}\varepsilon^2)$-sharpness of Proposition~\ref{prop:quant-trilemma} with binding distance $\varepsilon \propto v - 2\gamma$.}
\label{fig:inflation}
\end{figure}

\paragraph{Guardrails inflate reports exactly where they bind (Fig.~\ref{fig:inflation}).}
A soft cap on a bid-scaling proxy leaves within-range regret and report inflation at zero until the cap binds ($v > 2\gamma$), then both turn on, the field prediction of Corollary~\ref{cor:guardrails}. Of $60$ principals, the $44$ for whom the cap binds all inflate and all have $W_i > 0$.

The measured sweep traces the same profile on the models themselves. Pooling the seventeen models that share the rational bid, within-range regret and report inflation both rise with $v$ (Table~\ref{tab:per-value}), the empirical counterpart of the analytic curves above: the regret is quadratic in $v - 2\gamma$ (with $2\gamma = 80$) to the displayed precision through $v = 250$ (the grid discretizes the continuum values $0.117, 1.467, \dots$), the top point shaved by $0.12$ where the bid ceiling caps the frontier at $c(300) = 144$ ($14.40$ against the quadratic's $14.52$), and the inflation concentrates where the cap binds hardest, as Corollary~\ref{cor:guardrails} predicts.

\begin{table}[!ht]
\centering
\small
\caption{Per-value within-range regret and report inflation for the headline cluster, the seventeen models that share the rational bid, the empirical counterpart of the analytic profile in Figure~\ref{fig:inflation}. Both rise with the true value $v$.}
\label{tab:per-value}
\begin{tabular*}{\textwidth}{@{\extracolsep{\fill}} l r r r r r}
\toprule
true value $v$ & $100$ & $150$ & $200$ & $250$ & $300$ \\
\midrule
within-range regret $W_i$ & $0.12$ & $1.47$ & $4.32$ & $8.67$ & $14.40$ \\
report inflation & $1.25\times$ & $1.67\times$ & $1.88\times$ & $2.00\times$ & $2.00\times$ \\
\bottomrule
\end{tabular*}
\end{table}

\begin{figure}[tbp]
\centering
\begin{minipage}{0.49\textwidth}\centering
\begin{tikzpicture}
\begin{axis}[width=\linewidth, height=4.3cm,
  xmode=log, ymode=log, log basis x=2,
  xlabel={samples $N$}, ylabel={$|\hat g - g|$},
  tick label style={font=\footnotesize}, label style={font=\footnotesize},
  legend style={font=\tiny, at={(0.04,0.06)}, anchor=south west, draw=none, fill=none}]
\addplot[red!70!black, dashed, domain=16:2048, samples=2] {0.0558*sqrt(16/x)};
\addlegendentry{$1/\sqrt{N}$}
\addplot[blue!70!black, thick, mark=*, mark size=1.1pt] table[x=N, y=err] {figures/certification_decay.dat};
\addlegendentry{estimate error}
\end{axis}
\end{tikzpicture}
\\[2pt] {\footnotesize (a) certificate error}
\end{minipage}\hfill
\begin{minipage}{0.49\textwidth}\centering
\begin{tikzpicture}
\begin{axis}[width=\linewidth, height=4.3cm,
  xlabel={drift budget $B_T$}, ylabel={re-certifications},
  xmin=0, xmax=20, ymin=0, ymax=210,
  tick label style={font=\footnotesize}, label style={font=\footnotesize},
  legend style={font=\tiny, at={(0.04,0.96)}, anchor=north west, draw=none, fill=none}]
\addplot[red!70!black, dashed, thick] table[x=B, y=bound] {figures/certification_recert.dat};
\addlegendentry{Prop.~\ref{prop:online} bound}
\addplot[green!50!black, thick, mark=*, mark size=1pt] table[x=B, y=count] {figures/certification_recert.dat};
\addlegendentry{re-certifications}
\end{axis}
\end{tikzpicture}
\\[2pt] {\footnotesize (b) re-certification count}
\end{minipage}
\caption{(a) The sampling certificate's error against $N$, decaying at the $1/\sqrt{N}$ rate. (b) The re-certification count against the drift budget $B_T$, staying under the Proposition~\ref{prop:online} bound and scaling with $B_T$, not the number of updates.}
\label{fig:certification}
\end{figure}
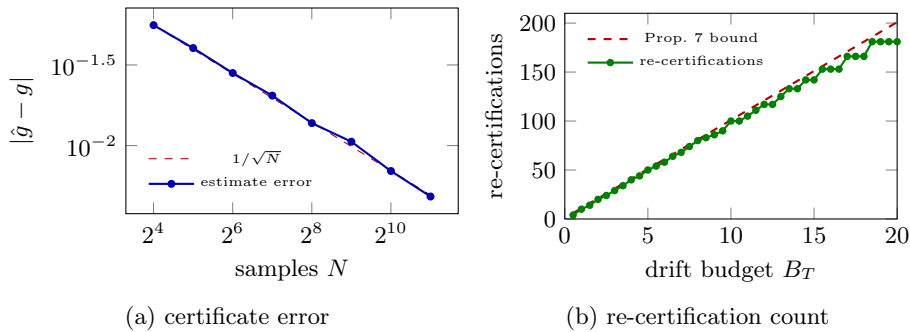

\paragraph{The certificate tracks the true gain and re-certifies under drift (Fig.~\ref{fig:certification}).}
The sampling certificate's estimate of the maximum manipulation gain decays toward the truth at the $1/\sqrt{N}$ rate, and under a stream of proxy updates the number of re-certifications stays under the Proposition~\ref{prop:online} bound and grows linearly in the total drift budget $B_T$, not the number of updates.

\section{Deferred proofs}
\label{app:proofs}

Each result of Sections~\ref{sec:erp}--\ref{sec:certification} is stated in the body and proved here, followed by the supporting and technical results.

\paragraph{Proposition~\ref{prop:inherit} (inheritance).}
There are two claims: that $\Gamma^{\pi}$ is interim $\varepsilon$-IC, and that truthful reporting reproduces the play of $\pi$. Fix an agent $i$, its true type $\theta_i$, and an arbitrary report $t_i \in \Theta_i$, and hold the opponents to truthful reporting, so they send the messages $\pi_{-i}$. The report is consumed only by the proxy, which converts $t_i$ into the message $\pi_i(t_i) \in M_i$; the mechanism then sees $\pi_i(t_i)$ against $\pi_{-i}$. Reporting $t_i$ yields interim utility $U_i(\pi_i(t_i), \pi_{-i} \mid \theta_i)$ and reporting truthfully yields $U_i(\pi_i(\theta_i), \pi_{-i} \mid \theta_i)$, abbreviated $U_i(\pi \mid \theta_i)$, and the gain is bounded in three steps:
\[
U_i(\pi_i(t_i), \pi_{-i} \mid \theta_i) - U_i(\pi \mid \theta_i)
\;\le\; \max_{m \in M_i} U_i(m, \pi_{-i} \mid \theta_i) - U_i(\pi \mid \theta_i)
\;\le\; R_i(\pi) \;\le\; \varepsilon.
\]
The first inequality holds because $\pi_i(t_i)$ is one message in $M_i$; the middle quantity is the interim regret at $\theta_i$, at most $R_i(\pi)$; the last is the $\varepsilon$-BNE hypothesis $\max_i R_i(\pi) \le \varepsilon$. As $i, \theta_i, t_i$ were arbitrary, $\Gamma^{\pi}$ is interim $\varepsilon$-IC. For the second claim, under truthful reporting each agent $j$ sends $\pi_j(\theta_j)$, so the outcome is $(g \circ \pi)(\theta)$, exactly the play of $\pi$ in $\Gamma$ at $\theta$. \qed

\paragraph{Lemma~\ref{lem:coarsen} (the input game is the coarsened base game).}
Fix $\pi$ and an agent $i$. A report strategy $\tau_i : \Theta_i \to \Theta_i$ composed with the proxy gives the base-game strategy $\pi_i \circ \tau_i : \Theta_i \to M_i$, whose range lies in $\Reach_i$ since each $\pi_i(\tau_i(\theta_i)) \in \pi_i(\Theta_i) = \Reach_i$. \emph{Onto:} for any base strategy $s_i$ with $s_i(\Theta_i) \subseteq \Reach_i$ and each $\theta_i$, the preimage $\pi_i^{-1}(s_i(\theta_i))$ is nonempty; choosing $\tau_i(\theta_i)$ in it gives $\pi_i \circ \tau_i = s_i$. \emph{Utility preservation:} at a report profile $\tau$ the wrapped mechanism plays $g\big((\pi_j(\tau_j(\theta_j)))_j\big)$, so $U_i^{\mathrm{input}}(\tau \mid \theta_i) = U_i\big((\pi_j \circ \tau_j)_j \mid \theta_i\big)$, which depends on $\tau_i$ only through $\pi_i \circ \tau_i$. Hence $\tau$ is an interim ($\varepsilon$-)equilibrium of the input game iff the composed profile is one of $\Gamma$ among the $\Reach$-valued strategies. \qed

\paragraph{Corollary~\ref{cor:guardrails} (guardrails).}
The guardrailed proxy $\pi'_i = c \circ \pi_i$ is itself a proxy, with reachable set $\Reach'_i = c(\Reach_i)$, so Theorem~\ref{thm:ttp} applies verbatim. By hypothesis $\pi'_i(\theta_i)$ does not maximize $U_i(\cdot\,, \pi_{-i} \mid \theta_i)$ over $\Reach'_i$, so $W_i(\theta_i; \pi') > 0$, and by Theorem~\ref{thm:ttp}(i) $G_i(\theta_i; \pi') = W_i(\theta_i; \pi') > 0$. \qed

\paragraph{Theorem~\ref{thm:ttp} (truth-to-proxy).}
All three claims rest on one observation: as the report $t_i$ ranges over $\Theta_i$, the message $\pi_i(t_i)$ ranges over exactly $\pi_i(\Theta_i) = \Reach_i$ and nothing else, so choosing a report is choosing a reachable action. \emph{(i)}~Maximizing utility over reports equals maximizing over $\Reach_i$, $\max_{t_i \in \Theta_i} U_i(\pi_i(t_i), \pi_{-i} \mid \theta_i) = \max_{m \in \Reach_i} U_i(m, \pi_{-i} \mid \theta_i)$; subtracting the honest payoff $U_i(\pi_i(\theta_i), \pi_{-i} \mid \theta_i)$ turns the two sides into $G_i(\theta_i;\pi)$ and $W_i(\theta_i;\pi)$, so $G_i = W_i$. \emph{(ii)}~Since $\Reach_i \subseteq M_i$, the maximum over $\Reach_i$ is at most that over $M_i$; subtracting the honest payoff, $W_i(\theta_i;\pi) \le \max_{m \in M_i} U_i(m, \pi_{-i} \mid \theta_i) - U_i(\pi_i(\theta_i), \pi_{-i} \mid \theta_i) \le R_i(\pi)$, the last step because $R_i(\pi)$ maximizes this term over types; with (i), $G_i \le R_i(\pi)$. \emph{(iii)}~Honest reporting is interim optimal iff $G_i(\theta_i;\pi) = 0$, iff $W_i(\theta_i;\pi) = 0$ by (i), which is range-optimality at $\theta_i$; hence $\Gamma^{\pi}$ is exactly IC iff $\pi$ is range-optimal at every type of every agent. \qed

\paragraph{Theorem~\ref{thm:trilemma} (the trilemma).}
Assume two of the three properties, truthfulness and capability-preservation, and derive the failure of the third; so suppose $c$ is both truthful and capability-preserving at $\theta_i$. Capability-preservation gives $a^\star \in \Reach'_i$. Because $\Reach'_i \subseteq \Reach_i$, because $a^\star$ already maximizes $U_i(\cdot\,, \pi_{-i} \mid \theta_i)$ over the larger set $\Reach_i$, and because $a^\star \in \Reach'_i$, the action $a^\star$ maximizes the same utility over the smaller set as well, so $\max_{m \in \Reach'_i} U_i(m, \pi_{-i} \mid \theta_i) = U_i(a^\star, \pi_{-i} \mid \theta_i)$. Truthfulness, $W_i(\theta_i; \pi') = 0$, says the guardrailed honest action $c(\pi_i(\theta_i))$ attains this constrained maximum, hence earns the same utility as $a^\star$; since $a^\star$ is the \emph{unique} maximizer over $\Reach_i \supseteq \Reach'_i$, the two coincide, $c(\pi_i(\theta_i)) = a^\star$, the negation of binding. Truthfulness and capability-preservation therefore force non-binding, and any two of the three preclude the third. \qed

\paragraph{Proposition~\ref{prop:quant-trilemma} (conservation of the binding loss).}
The within-range regret is $W_i(\theta_i;\pi') = \max_{m \in \Reach'_i} U_i(m, \pi_{-i} \mid \theta_i) - U_i(c(\pi_i(\theta_i)), \pi_{-i} \mid \theta_i)$, so the two losses telescope: $\kappa + W_i(\theta_i;\pi') = U^\star - U_i(c(\pi_i(\theta_i)), \pi_{-i} \mid \theta_i)$. The guardrailed honest action lies in $\Reach'_i \subseteq \Reach_i$ and, by $\varepsilon$-binding, at distance at least $\varepsilon$ from $a^\star$, so sharpness caps its value at $U^\star - \delta$. Hence $\kappa + W_i \ge \delta$. \qed

\paragraph{Lemma~\ref{lem:loyalty} (loyalty).}
At the true type $\theta_i$, loyalty makes the honest action $\pi_i(\theta_i)$ a maximizer of $U_i(\cdot\,, \pi_{-i} \mid \theta_i)$ over $C_i$, so honest reporting attains $\max_{m \in C_i} U_i(m, \pi_{-i} \mid \theta_i)$. The proxy produces only actions in $C_i$, so $\Reach_i \subseteq C_i$ and the within-range maximum over $\Reach_i$ cannot exceed this value. The gap defining $W_i(\theta_i; \pi)$ is therefore at most zero, and since $W_i$ is never negative it is exactly zero. The mechanism $g$ entered nowhere, so the conclusion holds for any $g$ under whose interim utilities the proxy is loyal in this sense. If every proxy is loyal over such a class, this gives $W_j = 0$ for all $j$, and $\Gamma^{\pi}$ is incentive-compatible by Theorem~\ref{thm:ttp}(iii). \qed

\paragraph{Proposition~\ref{prop:placement} (placement).}
Let $U' = \max_{m \in \Reach'_i} U_i(m, \pi_{-i} \mid \theta_i)$ be the constrained maximum; the guardrail is truthful iff its honest output attains $U'$ ($W_i = 0$) and capability-preserving iff $a^\star \in \Reach'_i$, in which case $a^\star$ is the unique maximizer over $\Reach'_i \subseteq \Reach_i$ and $U' = U_i(a^\star, \pi_{-i} \mid \theta_i)$. A non-binding guardrail has $c(\pi_i(\theta_i)) = a^\star \in \Reach'_i$, so it is capability-preserving, and its honest output attains $U'$, so $W_i = 0$: truthful. If the proxy re-optimizes over its constrained feasible set, its honest output maximizes over that set and hence over $\Reach'_i$ (Lemma~\ref{lem:loyalty}), so $W_i = 0$. If the output is instead clipped away from the reachable maximizer, its honest output does not attain $U'$, so $W_i > 0$; a clip is a fixed post-map, re-optimizing nothing, and in the output-clipping corner $a^\star \in \Reach'_i$ is the unique maximizer it misses. Capability is preserved exactly when $a^\star$ survives the constraint. Finally, a truthful and capability-preserving guardrail has its honest output maximize over $\Reach'_i$ while $a^\star \in \Reach'_i$ is the unique maximizer, so the honest output equals $a^\star$: unchanged, hence non-binding. No binding guardrail is both, which is Theorem~\ref{thm:trilemma}. \qed

\paragraph{Proposition~\ref{prop:neutral} (dominant-strategy neutrality).}
Fix $i$ and $\theta_i$. By hypothesis $a^\star_i(\theta_i)$ maximizes $U_i(\cdot\,, m_{-i} \mid \theta_i)$ over $\Reach_i$ for every $m_{-i} \in \Reach_{-i}$, and $a^\star_i(\theta_i) \in \Reach'_i \subseteq \Reach_i$; since a maximum over the larger set is attained inside the subset, $a^\star_i(\theta_i)$ also maximizes over $\Reach'_i$ against every $m_{-i} \in \Reach'_{-i} \subseteq \Reach_{-i}$, so it is weakly dominant in the guarded game as well. Both games therefore have $(a^\star_i(\theta_i))_i$ as a weakly-dominant-strategy equilibrium, unique in the unguarded game because $a^\star_i$ is; in the guarded game dominance is only required against $\Reach'_{-i}$, so ties may appear and uniqueness need not survive. Capability-preservation supplies reports $t_i, t'_i$ with $\pi_i(t_i) = \pi'_i(t'_i) = a^\star_i(\theta_i)$, so both equilibria send the identical message profile to $g$ at every type profile and induce the same allocation, payments, revenue, and welfare; the realizing reports differ, and for a monotone compressive cap the guarded report is the larger. \qed

\paragraph{Proposition~\ref{prop:destroy} (a capability-preserving guardrail can destroy welfare).}
Take the stag hunt of the body with two types per agent, a uniform prior, and a type-faithful proxy $\pi_i$ whose two type-images are $\mathsf{hare}$ and $\mathsf{stag}$, so $\Reach_i = \{\mathsf{hare}, \mathsf{stag}\}$; payoffs are type-independent. Against the honest opponent, an equal mix of $\mathsf{hare}$ and $\mathsf{stag}$, the reply $\mathsf{hare}$ earns $3$ and $\mathsf{stag}$ earns $2$, so $a^\star_i(\theta_i) = \mathsf{hare}$ at both types. The guardrail $c$ with $c(\mathsf{stag}) = c(\mathsf{hare}) = \mathsf{hare}$ is range-restricting, $\Reach'_i = \{\mathsf{hare}\}$, and capability-preserving, $a^\star_i = \mathsf{hare} \in \Reach'_i$. The report profile in which each agent sends the type reaching $\mathsf{stag}$ is an interim equilibrium of the unguarded input game, since $\mathsf{stag}$ is a best reply to $\mathsf{stag}$ ($4 > 3$); its outcome $(\mathsf{stag},\mathsf{stag})$ has welfare $8$. The guarded game reaches only $\mathsf{hare}$, so its sole outcome is $(\mathsf{hare},\mathsf{hare})$, welfare $6$, and no guarded equilibrium reproduces $(\mathsf{stag},\mathsf{stag})$. Restoring a third preserved message ($\Reach_i = \{\mathsf{hare}, \mathsf{alt}, \mathsf{stag}\}$ with $c$ deleting only $\mathsf{stag}$) leaves $\Reach'_i$ non-degenerate without changing the argument. \qed

\paragraph{Proposition~\ref{prop:hardness} (exact certification is intractable).}
Both reductions represent $g$ and the proxies as polynomial-size circuits, so $g \circ \pi$ evaluates in polynomial time and the hardness lies in counting and quantifying over types, not in the representation. \emph{(i)} Reduce from \textsc{Sat}. Given a formula $\phi$ over $x_1, \dots, x_{n-1}$, let agents $2, \dots, n$ have binary types $t_{-1} \in \{0,1\}^{n-1}$ (a truth assignment), agent $1$ have true type $\theta^0$ and alternative report $\theta^1$, outcomes $\{0,1\}$ with $u_1(1, \theta^0) = 1 > u_1(0, \theta^0) = 0$, each proxy the identity, and $g(t_1, t_{-1}) = \phi(t_{-1})$ if $t_1 = \theta^1$ and $0$ otherwise. All remaining utilities are constant, agent $1$'s at $\theta^1$ and every agent $j \ge 2$'s at all types, so truthful reporting is trivially dominant for them and the only agent-type that can violate IC is agent~$1$ at $\theta^0$. There, truthful reporting yields $0$ and $\theta^1$ yields $\phi(t_{-1})$, so agent~$1$ has a profitable misreport iff $\phi$ is satisfiable; thus $g \circ \pi$ is exactly dominant-strategy IC iff $\phi$ is unsatisfiable. Membership in coNP holds because a violating tuple is checked by two circuit evaluations. \emph{(ii)} With opponents' types uniform on $\{0,1\}^{n-1}$, the interim utility of $\theta^1$ at $\theta^0$ is $\E_{t_{-1}}[\phi(t_{-1})] = \#\{t_{-1} : \phi(t_{-1}) = 1\}/2^{n-1}$, so $G_1(\theta^0; \pi) = \#\phi / 2^{n-1}$; computing $G_1$ counts satisfying assignments, and $\#$\textsc{Sat} is $\#$P-complete. \qed

\paragraph{Lemma~\ref{lem:drift} (drift).}
Fix $i, \theta_i$ and a report $t_i$. The term $U_i(\pi_i(t_i), \pi_{-i} \mid \theta_i)$ is the expectation over $\theta_{-i}$ of $u_i$ at the message profile $(\pi_i(t_i), \pi_{-i}(\theta_{-i}))$; under the update each of its $n$ coordinates moves by at most $\Delta$, so by the $L$-Lipschitz hypothesis (applied at each $\theta_{-i}$ and averaged) the term shifts by at most $L n \Delta$. The gain $G_i(\theta_i; \cdot)$ is a maximum of such terms minus one such term; a maximum of functions each moving by $\le L n \Delta$ moves by $\le L n \Delta$, so by the triangle inequality $|G_i(\theta_i; \pi') - G_i(\theta_i; \pi)| \le 2 L n \Delta$. \qed

\paragraph{Proposition~\ref{prop:online} (online re-certification under a drift budget).}
Let $\mathrm{last}(t)$ be the epoch of the most recent fresh certificate, with bound $\skew{2}\hat\varepsilon_{\mathrm{last}}$ valid (probability $1-\delta$) at issuance. Telescoping Lemma~\ref{lem:drift} along $\pi^{\mathrm{last}(t)} \to \cdots \to \pi^t$ gives $\max_{i,\theta_i} G_i(\theta_i; \pi^t) \le \skew{2}\hat\varepsilon_{\mathrm{last}} + 2Ln\sum_{r=\mathrm{last}(t)}^{t-1}\Delta_r = \skew{2}\hat\varepsilon_{\mathrm{last}} + A_t$, with $A_t$ the drift accumulator (reset at each issuance). The policy re-estimates before $A_t$ exceeds $s$, so the live bound $\skew{2}\hat\varepsilon_{\mathrm{last}} + s$ dominates the true gain at every epoch, giving (i). For (ii), each re-certification after the first consumes $s$ of the total accumulator mass $2Ln\sum_{t<T}\Delta_t = 2Ln\,B_T$, giving at most $\lceil 2Ln\,B_T/s \rceil$ triggers plus the initial one; a union bound over the $K = \lceil 2Ln\,B_T/s \rceil + 1$ certificates gives confidence $1 - K\delta$. \qed


\end{document}